\documentclass{aa}
\usepackage[varg]{txfonts}

\usepackage{graphicx}
\usepackage{txfonts}
\usepackage{amsmath,amssymb}
\usepackage{soul}
\usepackage{xcolor}
\usepackage{enumitem}
\usepackage[colorlinks=true, allcolors=blue]{hyperref}
\usepackage{bbm}
\usepackage{xspace}

\definecolor{BrickRed}{RGB}{182, 50, 28}
\definecolor{forestgreen}{rgb}{0.13, 0.55, 0.13}

\newcommand{\citeg}[1]{\citep[e.g.,][]{#1}}

\newcommand{\Yo}{\texttt{YOLO}\xspace}
\newcommand{\YoCL}{\texttt{YOLO-CL}\xspace}
\newcommand{\YoVtwo}{\texttt{YOLO9000}\xspace}
\newcommand{\YoV}{\texttt{YOLOv3}\xspace}

\begin{document}

\title{YOLO-CL: Galaxy cluster detection in the SDSS with deep machine learning}
\titlerunning{Galaxy cluster detection in the SDSS with \YoCL}
\author{
   Kirill Grishin\inst{1},
   Simona Mei\inst{1,2},
   Stéphane Ili\'c \inst{3,1,4,5}
}

 \institute{Universit\'e Paris Cit\'e, CNRS(/IN2P3), Astroparticule et Cosmologie, F-75013 Paris, France \email{grishin@apc.in2p3.fr, mei@apc.in2p3.fr}
\and Jet Propulsion Laboratory and Cahill Center for Astronomy \& Astrophysics, California Institute of Technology, 4800 Oak Grove Drive, Pasadena, California 91011, USA
\and Universit\'e PSL, Observatoire de Paris, Sorbonne Universit\'e, CNRS, LERMA, F-75014, Paris, France
\and CNES Centre National d'\'Etudes Spatiales, Toulouse, France
\and  IJCLab, Université Paris-Saclay, CNRS/IN2P3, IJCLab, 91405 Orsay, France \email{ilic@ijclab.in2p3.fr}
   }
   
\date{}

\abstract{Galaxy clusters are a powerful probe of cosmological models. Next generation large-scale optical and infrared surveys will reach unprecedented depths over large areas and require highly complete and pure cluster catalogs, with a well-defined selection function. We have developed a new cluster detection algorithm \YoCL, which is a modified version of the state-of-the-art object detection deep convolutional network \Yo, optimized for the detection of galaxy clusters. We trained \YoCL on color images of the redMaPPer cluster detections in the SDSS. We find that \YoCL detects $ 95-98\%$ of the redMaPPer clusters, with a purity of $95-98\%$ calculated by applying the network to SDSS blank fields. When compared to the MCXC2021 X-ray catalog in the SDSS footprint, \YoCL recovers all clusters at $L_X \gtrsim 2-3 \times 10^{44}$~erg/s, $M_{500} \gtrsim 2-3 \times 10^{14} M_{\odot}$, $R_{500} \gtrsim 0.75-0.8$~Mpc and $0.4 \lesssim z \lesssim 0.6$. When compared to the redMaPPer detection of the same MCXC2021 clusters, \YoCL is more complete than redMaPPer, which means that the neural network improved the cluster detection efficiency of the redMaPPer algorithm. In fact, we find that \YoCL  detects $\sim 98\%$ of the MCXC2021 clusters with mean X-ray surface brightness $ {\rm I_{X, 500}} \gtrsim 20 \times 10^{-15} \ {\rm erg/s/cm^2/arcmin^2}$  at $0.2 \lesssim z \lesssim 0.6$  and $\sim 100\%$ of the MCXC2021 clusters with  $ {\rm I_{X, 500}} \gtrsim 30 \times 10^{-15} \ {\rm erg/s/cm^2/arcmin^2}$ at $ 0.3 \lesssim z \lesssim 0.6$, while 
redMaPPer detects $\sim 98\%$ of the MCXC2021 clusters with $ {\rm I_{X, 500}} \gtrsim 55 \times 10^{-15} \ {\rm erg/s/cm^2/arcmin^2}$ at $0.2 \lesssim z \lesssim 0.6$ and $\sim 100\%$ of the MCXC2021 clusters with $ {\rm I_{X, 500}} \gtrsim 20 \times 10^{-15} \ {\rm erg/s/cm^2/arcmin^2}$ at  $0.5 \lesssim z \lesssim 0.6$.  The \YoCL selection function is approximately constant with redshift, with respect to the MCXC2021 cluster X-ray surface brightness. \YoCL shows high performance when compared to traditional detection algorithms applied to SDSS. Deep learning networks  benefit from a strong advantage over traditional galaxy cluster detection techniques because they do not need galaxy photometric and photometric redshift catalogs. This eliminates systematic uncertainties that can be introduced during source detection, and photometry  and photometric redshift measurements. Our results show that \YoCL is an efficient alternative to traditional cluster detection methods. In general, this work shows that it is worth exploring the performance of deep convolution networks for future cosmological cluster surveys, such as the Rubin/LSST, Euclid or the Roman Space Telescope surveys. 
}


\maketitle

\section{Introduction}
\label{sec:Intro}

Clusters of galaxies are powerful probes to constrain cosmological models. In fact, since they are the largest and most massive gravitationally bound systems in the Universe, their abundance probes the growth history of structures \citep[e.g.,][]{2011ARA&A..49..409A}. 
Future large-scale surveys, such as the Dark Energy Survey\footnote{\url{darkenergysurvey.org}} \citep{2018ApJS..239...18A}, the Dark Energy Spectroscopic Instrument\footnote{\url{desi.lbl.gov}} \citep{2019AJ....157..168D}, the Vera C. Rubin Observatory\footnote{\url{vro.org}} \citep[formerly Large Synoptic Survey Telescope,][]{2018cosp...42E1651K}, or the \textit{Euclid} satellite\footnote{\url{euclid-ec.org}} \citep{Euclid-r} and the Nancy Grace Roman Space Telescope \citep{2021MNRAS.507.1746E}, will use large cluster samples as cosmological probes, and need the development of fast and efficient cluster detection algorithms. These  surveys will reach unprecedented depths over large areas and will need highly complete and pure cluster catalogs, with a well-defined selection function.

Cluster detection algorithms have been developed by the astronomical community at different wavelengths.  The detection in optical and near-infrared bandpasses is mainly based on the search of spatial overdensities of a given class (quiescent, line-emitters, massive, etc.) of galaxies  \citeg{2005ApJS..157....1G, 2009ApJ...697.1842K, 2012ApJ...746..188M, 2010MNRAS.404.1551S, 2011MNRAS.413.2883B, 2014ApJ...785..104R, 2013ApJ...769...79W, 2014ApJ...786...17W},
while detections in the X-rays and submillimeter rely on the assumption of model profiles that fit the data as for example a characteristic galaxy cluster luminosity and a radial profile  \citeg{2007A&A...461...81O, 2009A&A...494..845G, 2016A&A...594A..27P, 2004A&A...425..367B, 2011ApJ...737...61M, 2021A&A...647A...1P}. 

\begin{figure*}[ht!]
    \includegraphics[width=2\columnwidth]{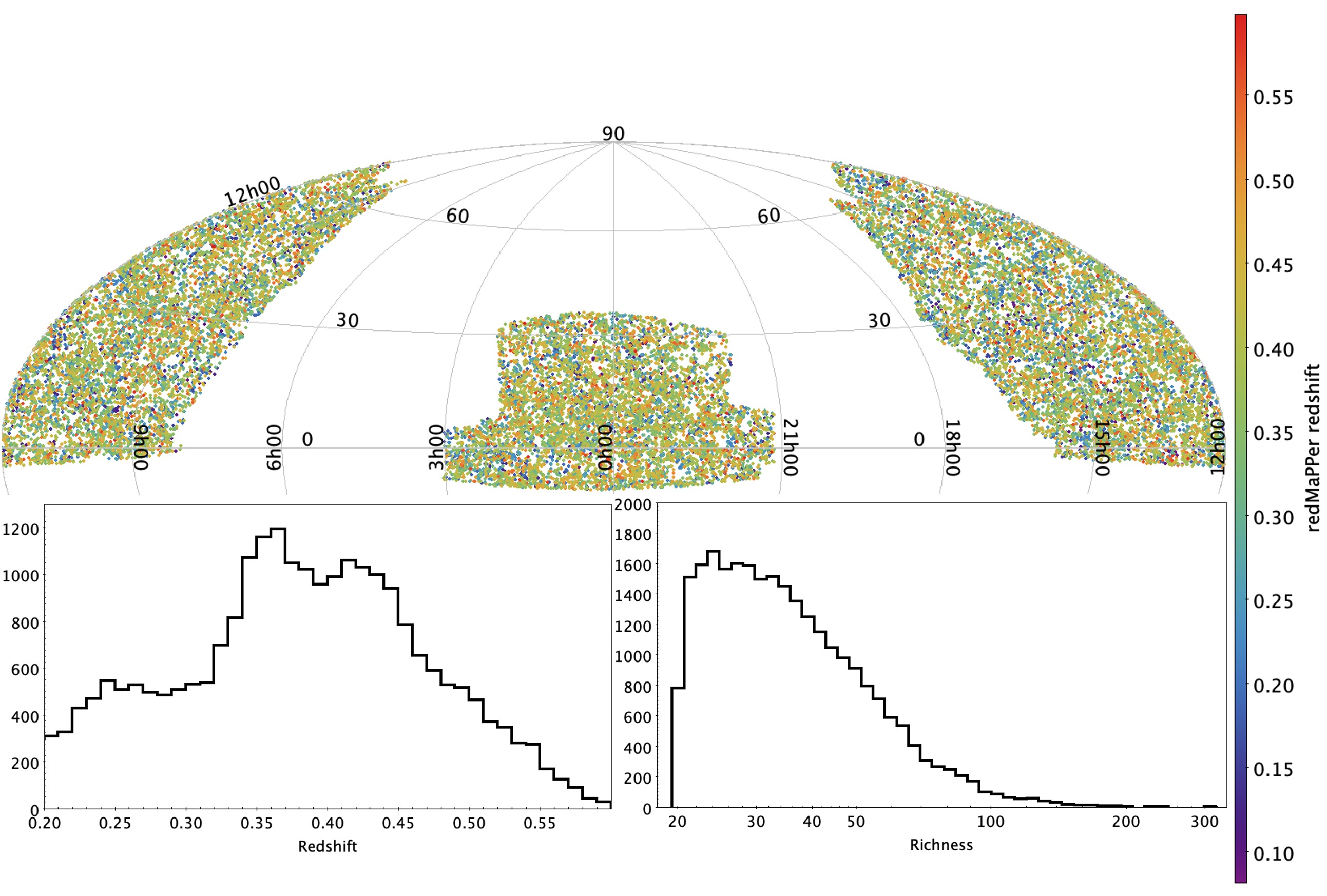}
    \caption{The redMaPPer sample of 24,406 clusters used to train and validate our network. Top: Sky map of the positions of the redMaPPer clusters in celestial coordinates, where color indicates the photometric redshift of the cluster as estimated by the redMaPPer algorithm. Bottom: the training and validation redMaPPer sample redshift (left) and richness (right) distribution.}
    \label{fig:map_redmap}
\end{figure*}

In the local Universe and up to $z\sim1$, the reference cluster catalog has been the all-sky ROSAT X-ray catalog \citep{1998MNRAS.301..881E,1999A&A...349..389V,2004A&A...425..367B} for a couple of decades. In the past ten years though, optical and millimeter-wave surveys provided large cluster samples that have been used to constrain our cosmological model parameters 
\citeg{2010ApJ...708..645R, 2011ARA&A..49..409A,  2013JCAP...07..008H, 2015ApJS..216...27B, 2015A&A...582A..29P, 2016ApJ...832...95D, 2021PhRvD.103d3522C, 2022arXiv220712429C}. 
 
The  X-ray survey eROSITA \citep{2012arXiv1209.3114M},
the next generation CMB surveys (e.g. Simons Observatory \citealp{2019JCAP...02..056A}), and the Euclid \citeg{2015MNRAS.452..549A, 2019A&A...627A..23E}, the Nancy Grace Roman Space Telescope \citep{2021MNRAS.507.1746E} and Rubin Observatory Legacy Survey of Space and Time (LSST)\footnote{\url{https://lsst.slac.stanford.edu/}} cluster surveys \citeg{2019ApJ...873..111I} will extend these cluster samples at lower mass and higher redshifts. In the high redshift Universe ($z \gtrsim 1.5$), cluster detection will be mainly performed by optical and infrared surveys, combined with radio and far-infrared observations. In fact, clusters are predicted to be less massive \citep{2013ApJ...779..127C}, and the X-ray and SZ (Sunyaev-Zel'dovich) signal is fainter \citeg{2017MNRAS.464.2270A}, limiting the performance of X-ray and SZ cluster detection algorithms.

The use of Deep Machine Learning (ML) algorithms in various areas of astrophysics has been rising for the past decade, with applications ranging from the analysis of galaxy surveys \citep[see e.g.][for a recent review]{2022arXiv221001813H}, photometric redshift estimations \citep[see e.g.][for a recent review]{2021MNRAS.505.4847H} to dark matter map reconstructions \citep[e.g.][]{2020MNRAS.492.5023J}. In particular, neural networks (CNN) have proved especially useful in object detection and characterization \citeg{2015ApJS..221....8H,2018ApJ...858..114H,2018MNRAS.478.5410D, 2019A&A...621A..26P,2021MNRAS.501.4359Z,2022A&A...657A..90E,2022A&A...665A..34D, 2022arXiv220614944E, 2022arXiv220913074E}, and in particular in galaxy cluster detection \citeg{2019MNRAS.490.5770C,2020A&A...634A..81B,2021A&A...653A.106H, 2021MNRAS.507.4149L}.

Many object detection algorithm have been developed in the field of Deep ML (see the recent reviews of \citealt{2019arXiv190505055Z} and \citealt{2021arXiv210411892Z}), most of which have not been applied in the field of astrophysics. In this paper, we use as a basis the architecture of the well-known detection-oriented deep ML neural network ``\texttt{You only look once}" \citep[\Yo,][]{2015arXiv150602640R,2016arXiv161208242R} to detect clusters of galaxies in the Sloan Digital Sky Survey, and assess its efficiency.  The \Yo algorithm has been developed for a very wide range of real-life situations, for example for face detection, for the analysis of medical images, and for self-driving cars. Its last implementations, among which \YoV developed by \citet{2018arXiv180402767R}, are particularly efficient for multiple object detection and well-adapted to cluster detection.

Our results show that our \Yo network, which we called \YoCL (see Section~\ref{sec:yolo}), adapted for the detection of galaxy clusters, shows a high performance with respect to traditional cluster detection algorithms in obtaining dependable cluster catalogs with high levels of completeness and purity. Our results show that our \YoCL cluster catalogs have a purity of $95-98\%$ on blank fields and a completeness of $\sim 98\%$ for X-ray detected clusters with $ {\rm I_{X, 500}} \gtrsim 20 \times 10^{-15} \ {\rm erg/s/cm^2/arcmin^2}$  at $0.2 \lesssim z \lesssim 0.6$ , and of $\sim 100\%$ for clusters with $ {\rm I_{X, 500}} \gtrsim 30 \times 10^{-15} \ {\rm erg/s/cm^2/arcmin^2}$ at $ 0.3 \lesssim z \lesssim  0.6$. Our selection function is flat as a function of redshift, when considering mean X-ray surface brightness.

In Section~\ref{sec:obs}, we describe the data and the catalog used for the training and validation of our network. Section~\ref{sec:yolo} presents our network implementation and how we build our cluster catalog. In Section~\ref{sec:red}, we compare our results with the training cluster catalog and a X-ray cluster catalog. We discuss and summarize our results in Section~\ref{sec:discussion} and Section~\ref{sec:summary}, respectively.

\section{Observational dataset} \label{sec:obs}

For the past two decades, the Sloan Digital Sky Survey (SDSS\footnote{\url{https://classic.sdss.org/}}) has been the largest imaging and spectroscopic survey of the local Universe \citep{2000AJ....120.1579Y}. It uses a dedicated 2.5-m wide field-of-view optical telescope, located at the Apache Point Observatory, and has provided astronomers with a tremendous amount of data.
This wealth of data has consistently yielded cosmological constraints via the various SDSS Data Releases (DR), so far culminating in the 17th data release \citep[DR17,][]{2022ApJS..259...35A}.

To train and test the application of \Yo to cluster detection, we focus on the 
 most complete and pure SDSS cluster catalog (see also Section~\ref{sec:discussion}), the redMaPPer DR8 (Data Release 8) catalog from \citet{2014ApJ...785..104R}. The redMaPPer algorithm is a red sequence cluster finder specifically designed for  large photometric surveys. The redMaPPer algorithm was applied to the $\sim$10,000 square degrees of the SDSS DR8 data release, yielding a catalog\footnote{Version 6.3 of the catalog, from \url{risa.stanford.edu/redMaPPer}.} of 26,111 clusters over the redshift range z $\in$ [0.08, 0.55]. With respect to the MCXC X-ray detection catalog \citep{2011A&A...534A.109P}, the redMaPPer catalog was found to be 100\% complete up to $z=0.35$, above the X-ray temperature $T_X \gtrsim 3.5 keV$, and $L_X \gtrsim 2 \times 10^{44}$ erg  s$^{-1}$, decreasing to 90\% completeness at $L_X \sim 10^{43}$ erg \ s$^{-1}$. 86\% of the redMaPPer clusters are correctly centered with respect to their X-ray centers \citep{2014ApJ...785..104R}. All redMaPPer rich clusters ($\lambda > 100$) are detected in the X-ray ROSAT All Sky Survey \citep{1999A&A...349..389V}.
  
 To train and validate our network, we excluded clusters with redshifts $z<0.2$ which cover regions in the sky larger than the images that we consider, and worked with a final sample of 24,406 clusters, whose distribution is shown in Figure~\ref{fig:map_redmap}. 
 For each cluster, the algorithm provides its position,  the richness $\lambda$\footnote{The cluster richness is defined as the number of cluster members above a given luminosity. For redMaPPer it is defined as a sum of the probability of being a cluster member over all galaxies in a cluster field~\citep{2009ApJ...703..601R}.} as a proxy for cluster mass, and a list of cluster members \citep{2014ApJ...783...80R}.

\begin{figure}[t!]
    \centering
    \includegraphics[width=\columnwidth]{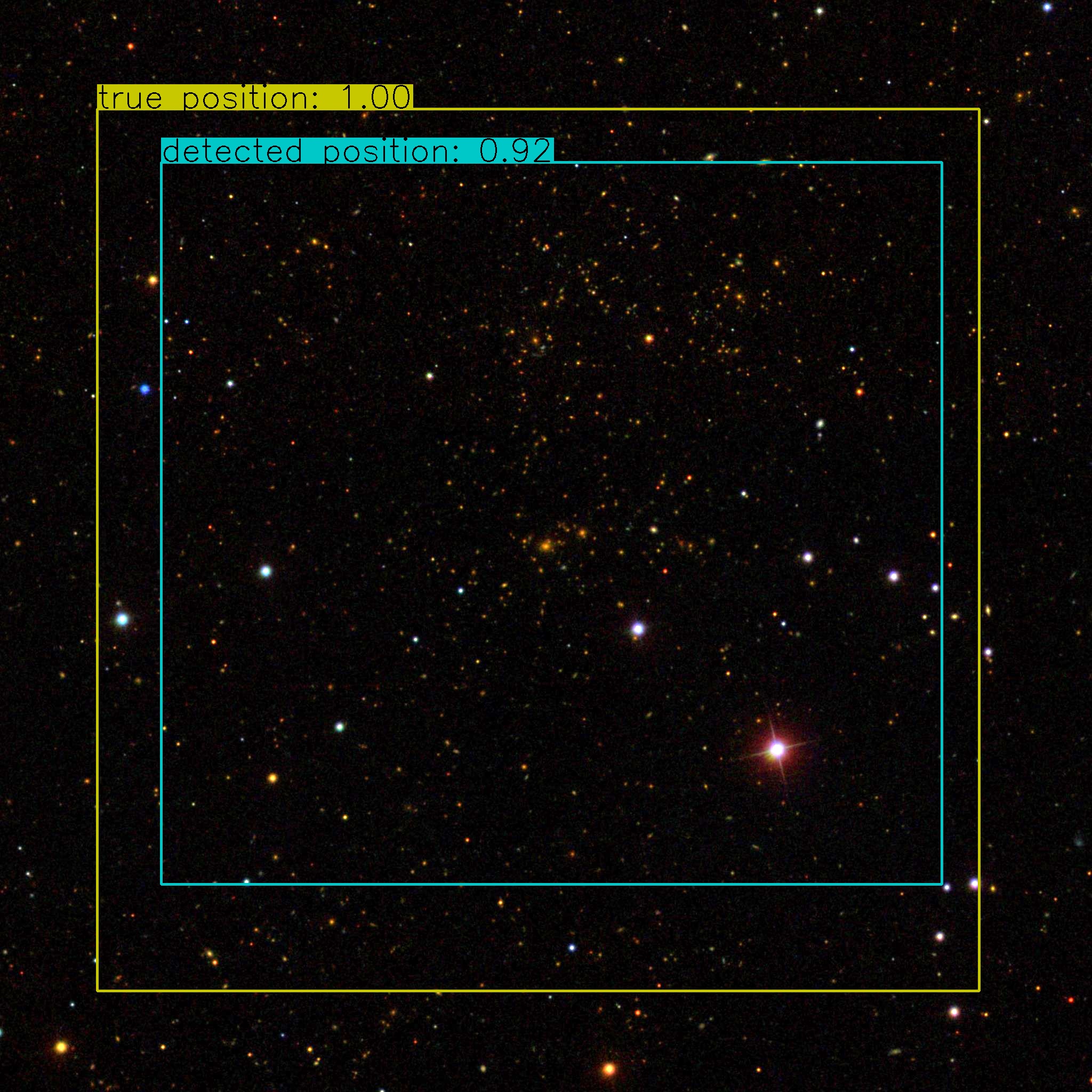}
    \caption{SDSS image cutout of a redMaPPer cluster in our sample. The yellow box corresponds to the minimal rectangle encompassing all redMaPPer cluster members, which is the box used to train \YoCL. In cyan, the box detected by our network \YoCL, with the associated confidence level in the top left corner. The image size is 13.5 x 13.5 arcmin$^2$, and the pixel size is 0.396 arcsec. }
    \label{fig:example_detect}
\end{figure}

For the network training and validation, we retrieve JPEG versions of the original SDSS DR16 raw images for each redMaPPer cluster, and constructed color images using the \texttt{ImgCutout} web service\footnote{\url{http://skyserver.sdss.org/dr16/en/help/docs/api.aspx\#imgcutout}}
 by querying the SDSS Catalog Archive Server databases. These images are based on the \textit{g}, \textit{r}, and \textit{i}-band FITS corrected frame files from the Science Archive Server, and the color images are built using the conversion algorithm\footnote{Detailed here: \url{https://www.sdss.org/dr16/imaging/jpg-images-on-skyserver}}
based on \citet{2004PASP..116..133L}.  These three bandpasses are sufficient to identify passive early-type galaxies in clusters at $z<1$.  Figure~\ref{fig:example_detect} shows an example of such cutout images.

\section{ \YoCL : our \Yo network for galaxy cluster detections} \label{sec:yolo}

\subsection{The \Yo network}

The \Yo \citep{2015arXiv150602640R} network is a state-of-the-art, real-time object detection deep convolutional network.

Competing architectures in the ML literature tend to apply first a ``localizer'' network on a given image, at multiple locations and at multiple scales, and assign a detection probability. The high probability regions of the image are considered as detections, and are then classified using a separate network.

The \Yo architecture uses a different approach: it applies a single neural network to the full image, combining the detection and classification into a single process. This gives the network several advantages over classifier-based systems, because its predictions take into account the global context of the image. It also has the advantage of making predictions with a single network evaluation, unlike systems such as R-CNN \citep[Region Based Convolutional Neural Networks,][and following iterations Fast and Faster R-CNN]{2013arXiv1311.2524G} which require thousands of evaluations for a single image. This can result in several orders of magnitude faster \Yo execution times, compared to R-CNN and Fast R-CNN.

In practice, the network divides the image into a $S{\times}\,S$ grid of regions (or cells), within which detection and classification are performed. \Yo predicts $B$ bounding boxes\footnote{For the sake of completeness, we note here that the \Yo network does not actually predicts the positions and dimensions of the bounding boxes directly, but rather offsets from a fixed set of $B$ boxes called ``anchors'' which act as priors to facilitate the training of the network. Those anchor boxes are usually derived from the training set by running a $k$-means clustering algorithm (with $k=B$) on the set of true bounding boxes.} per region, with their associated ``objectness'' probability (i.e. how confident we are that there is an object in the box) and ``class probabilities" (i.e. for a set of classes, the respective probabilities that the potential object belongs to them). Both $B$ and $S$ are hyperparameters of the network and can be adjusted by the user.

The predicted bounding boxes and their associated probabilities are returned by the network in the following format:
\begin{equation}
    \label{eq:bbox_vec}
    (x, y, w, h, C, p(c_1), \dots, p(c_n))
\end{equation}
where $(x,y)$ are the coordinates of the box center, $w$ and $h$ its width and height, $C$ the objectness, and $p(c_1), \dots, p(c_n)$ the probabilities (summing to one) that the object in the box belongs respectively to the class $c_1, \dots, c_n$.

When training a \Yo network using a set of images (with their associated ``true'' bounding boxes), we optimise the following multi-part loss function $\mathcal{L}$ \citep{2015arXiv150602640R}:
\begin{equation}
    \label{eq:yolo_loss}
    \mathcal{L} = \mathcal{L}_{\rm bbox} + \mathcal{L}_{\rm obj} + \mathcal{L}_{\rm class} \ .
\end{equation}
The first term of Eq.~(\ref{eq:yolo_loss}) is the ``bounding box loss'':
\begin{multline}
    \label{eq:bbox_loss}
    \mathcal{L}_{\rm bbox} = \alpha_\textbf{coord}
    \sum_{i = 0}^{S^2}
        \sum_{j = 0}^{B}
         {\mathbbm{1}}_{ij}^{\text{obj}}
                \left[
                \left(
                    x_i - \hat{x}_i
                \right)^2 +
                \left(
                    y_i - \hat{y}_i
                \right)^2
                \right]
    \\
    + \alpha_\textbf{coord}
    \sum_{i = 0}^{S^2}
        \sum_{j = 0}^{B}
             {\mathbbm{1}}_{ij}^{\text{obj}}
             \left[
            \left(
                \sqrt{w_i} - \sqrt{\hat{w}_i}
            \right)^2 +
            \left(
                \sqrt{h_i} - \sqrt{\hat{h}_i}
            \right)^2
            \right]
\end{multline}
where the ({\it x, y}) coordinates represent the center of the box relative to the bounds of the grid cell, and $w$ and $h$ are the width and height of the box. The symbol $\mathbbm{1}_i^{\text{obj}}$ denotes if an object appears in cell $i$ and $\mathbbm{1}_{ij}^{\text{obj}}$ denotes that the $j$th bounding box predictor in cell $i$ is ``responsible'' for that prediction. In these equations, and those below, the variables with a hat over them are the "true values" that the network is learning.

The second term is the ``objectness loss'':
\begin{equation}
     \mathcal{L}_{\rm obj} = \sum_{i = 0}^{S^2}
        \sum_{j = 0}^{B}
            {\mathbbm{1}}_{ij}^{\text{obj}}
            \left(
                C_i - \hat{C}_i
            \right)^2
    + \alpha_\textrm{noobj}
    \sum_{i = 0}^{S^2}
        \sum_{j = 0}^{B}
        {\mathbbm{1}}_{ij}^{\text{noobj}}
            \left(
                C_i - \hat{C}_i
            \right)^2
\end{equation}
where $C$ represents the conditional class probability. Finally, the last term represents the ``classification loss'':
\begin{equation}
    \label{eq:class_loss}
    \mathcal{L}_{\rm class} = \sum_{i = 0}^{S^2}
    {\mathbbm{1}}_i^{\text{obj}}
        \sum_{c \in \textrm{classes}}
            \left(
                p_i(c) - \hat{p}_i(c)
            \right)^2
\end{equation}
where the $p_i(c)$ correspond to the probabilities to belong to a certain class $i$. The $\alpha_{coord}$ and the  $\alpha_{noobj}$ coefficients appearing in the previous formulas can be changed to give more weight to certain components of the total loss. We choose to set $\alpha_{coord}= \alpha_{noobj}=1$

The first version  of the \Yo network used a Darknet-19 neural network architecture, which contains 19 layers, as the feature extractor. Its second version (\YoVtwo, \citealt{2016arXiv161208242R}) added 11 more layers to Darknet-19, reaching a total of 30. These first architectures had difficulties to detect small objects due to the coarseness of their $S{\times}\,S$ grid.

We base our cluster detection network on the third iteration of \Yo \citep{2018arXiv180402767R}, \YoV, which represents a significant improvement over the first two versions, and, while several other \Yo versions have been developed afterwards, we will consider their application only in future work. The feature extractor was replaced by Darknet-53 and residual networks. The new extractor uses 53 convolution layers, with consecutive 3x3 and 1x1 convolution layers followed by a skip connection (introduced by ResNet, \citealt{2015arXiv151203385H}) to help the activations to propagate through deeper layers without gradient diminishing. With 53 additional layers for the detection, \YoV totals 106 fully convolutional layers. Its larger size makes it slower as compared to previous iterations, but significantly enhances its accuracy.

Moreover, \YoV introduces a multi-scale feature in the detection process. In practice, instead of producing a single ``feature map'' (to be fed to the detection part of the network) at a single $S{\times}\,S$ resolution, Darknet-53 produces three different maps at three different levels of resolution. The underlying idea here is to provide the detection network with feature maps at three different scales: the coarser the map, the bigger the objects it would detect.

\begin{table*}[ht!]
   \centering
    \caption{Settings used for the \YoCL training}
    \resizebox{!}{1cm}{
    \begin{tabular}{ c c c c c c }
    \hline
    Image resolution & Batch size & Number of training & Data augmentation & Augmentation frequency & gIoU threshold\\
     &  & epochs & technique &   per technique& \\[0.1cm]
    \hline
    \rule{0pt}{\normalbaselineskip} 
    $1024\times1024$ & 2 & 100 & horizontal flip, vertical flip, transpose & 50 \% & 50 \% \\[0.1cm]
    $512\times512$ & 8 & 100 & horizontal flip, vertical flip, transpose & 50 \% & 50 \% \\
    \hline
    \end{tabular}}
    \label{tab:settings}
\end{table*}

\Yo networks have been used in applications in astrophysics to detect galaxies and other sources \citep{2018A&C....25..103G,2021MNRAS.508.2039H}, and astrophysical transients \citep{2022AJ....164..250L}.

\subsection{Network optimization for galaxy cluster detection}

\subsubsection{Modifications to the original \YoV network} 
To optimize \YoV for galaxy cluster detection, we applied several modifications to its standard architecture, and we call our new implementation \YoCL (\Yo for CLuster detection). \YoCL is based on a TensorFlow implementation of the \YoV architecture\footnote{\url{https://github.com/YunYang1994/TensorFlow2.0-Examples}}. 
The network was trained on a NVIDIA Tesla V100-SXM2-32GB GPU, equipped with 32 GB of memory. 

Our first modification is the definition of a single (instead of multiple classes in the original network) object class, clusters, and  we therefore removed the $\mathcal{L}_{\rm class}$ term, defined in Eq.~(\ref{eq:class_loss}), from the original loss function of Eq.~(\ref{eq:yolo_loss}). This results in fewer network weights, which leads to a lighter, faster, and easier-to-train network.

\begin{figure*}[ht!]
    \centering
    \includegraphics[width=0.48\textwidth]{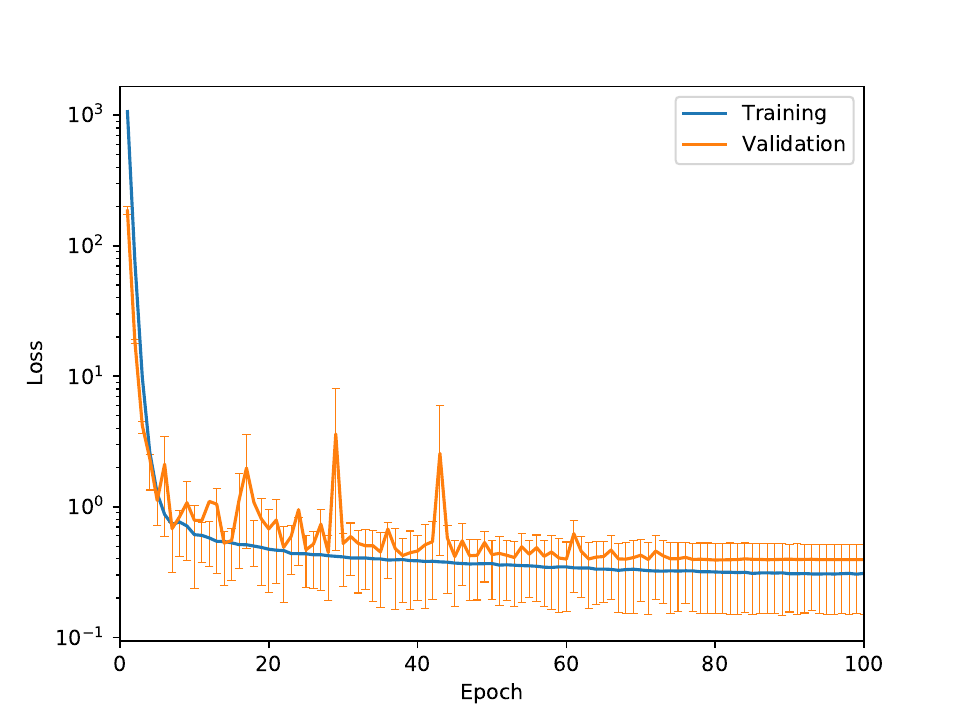}
    \includegraphics[width=0.48\textwidth]{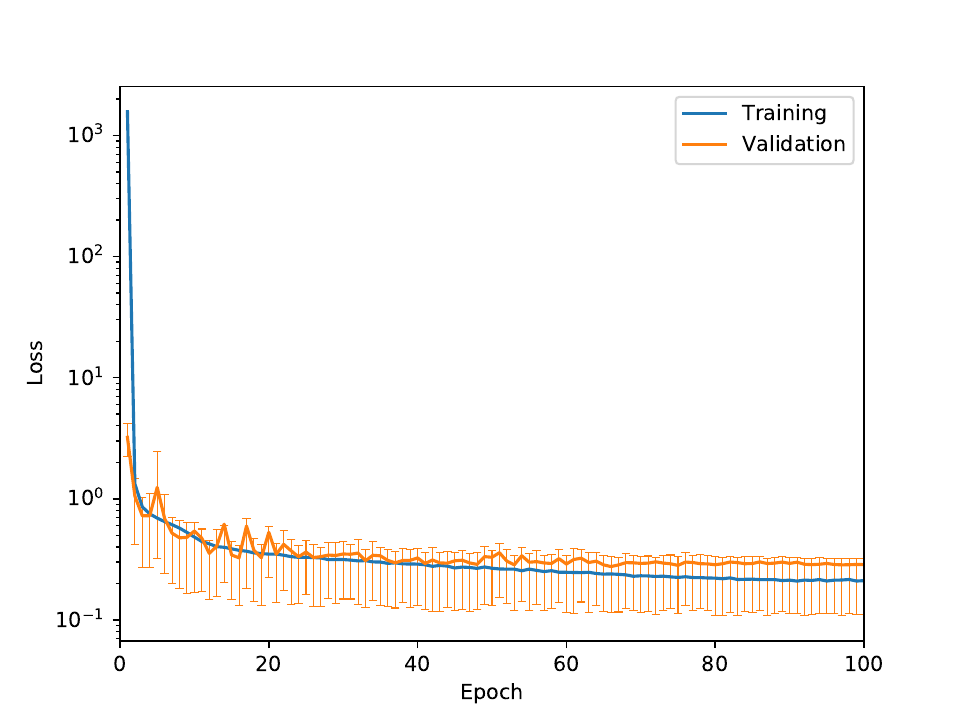}
    \caption{The mean training (blue) and validation (orange) loss for \YoCL when using 512$\times$512 (left), and 1024$\times$1024 (right) resampled SDSS images. The vertical bars show the 1~$\sigma$ standard deviation of the validation loss. The training and validation loss functions converge in a smooth way. The good agreement between training and validation loss excludes significant overfitting and confirming the network stability in both cases.} \label{fig:loss}
\end{figure*}

\begin{figure*}
    \centering
    \includegraphics[width=0.4\textwidth]{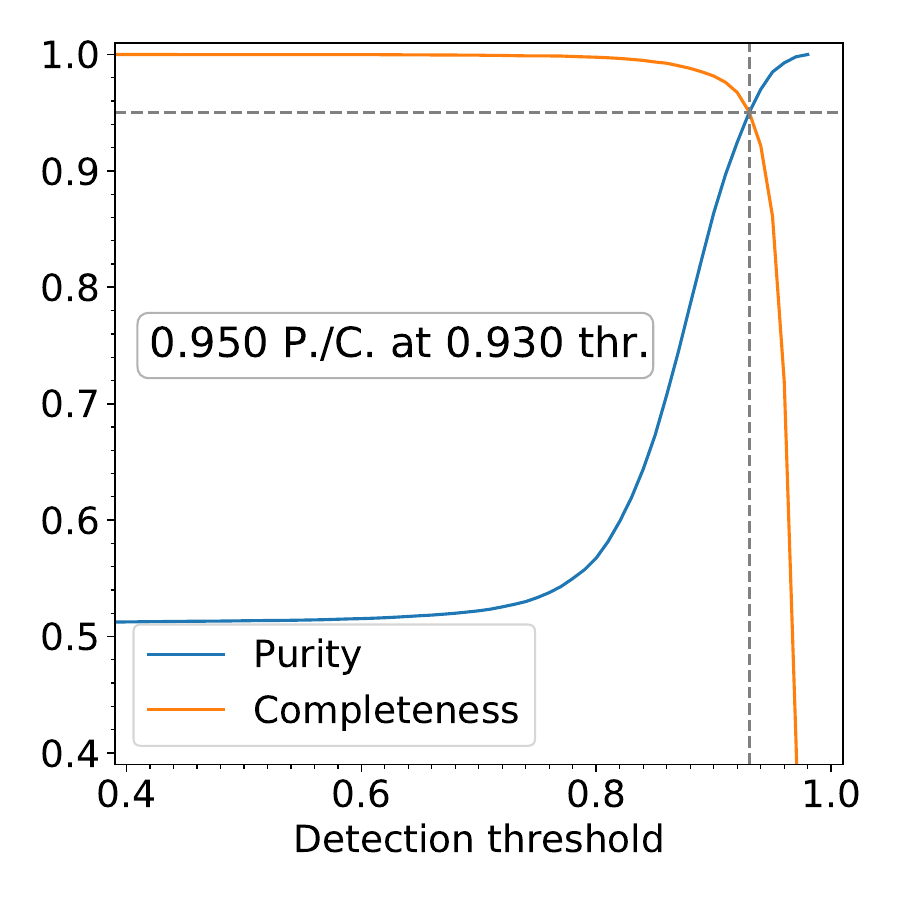}
     \includegraphics[width=0.4\textwidth]{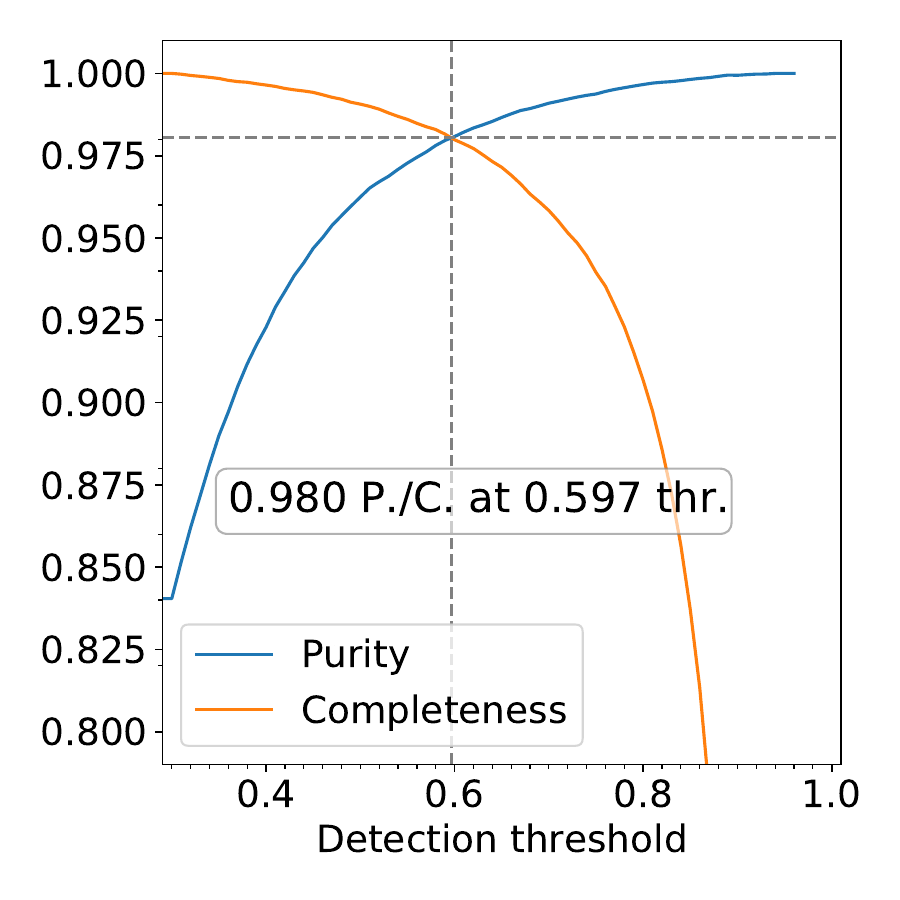}
    \caption{The \YoCL cluster catalog completeness and purity as a function of the detection threshold when using 512$\times$512 (left), and 1024$\times$1024 (right) resampled SDSS images. When using the 512x512 and the 1024x1024 resampled SDSS images, we obtain a completeness and purity of 95\% and 98\%, respectively, at the optimal (see text) threshold of 93\% and 60\%, respectively. Overall, \YoCL has a very good performance in the redshift range covered by redMaPPer. } \label{fig:cp}
\end{figure*}

 \begin{figure*}
    \centering
    \includegraphics[width=0.35\textwidth]{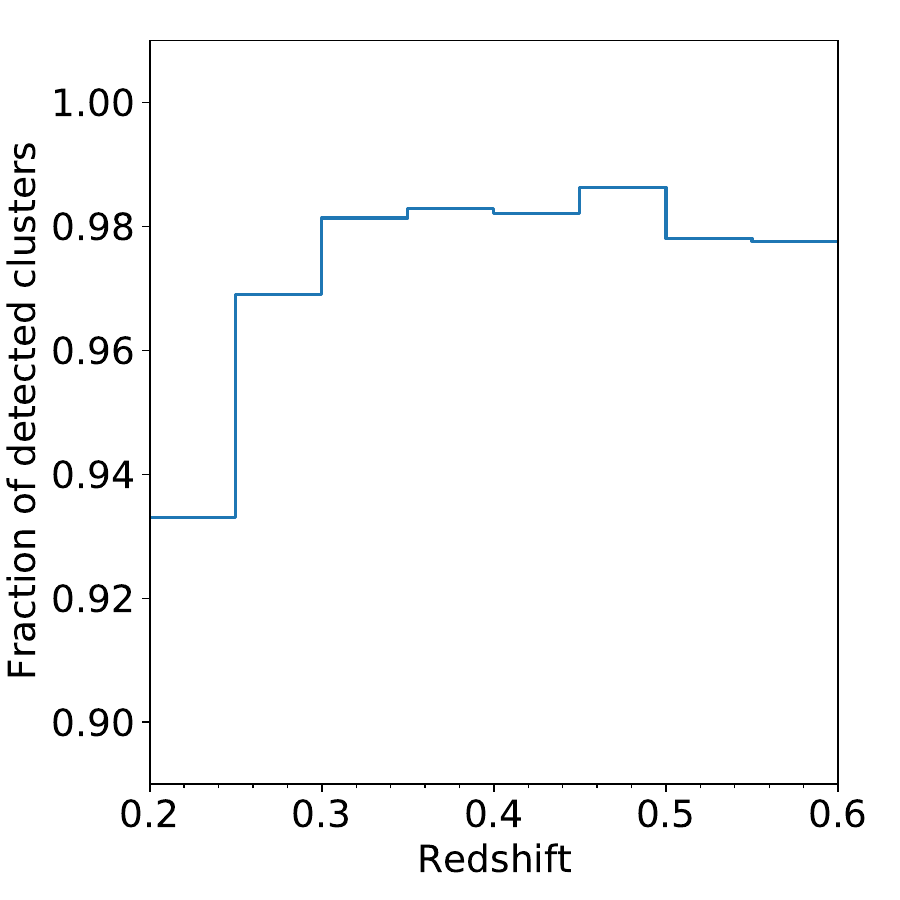}
     \includegraphics[width=0.35\textwidth]{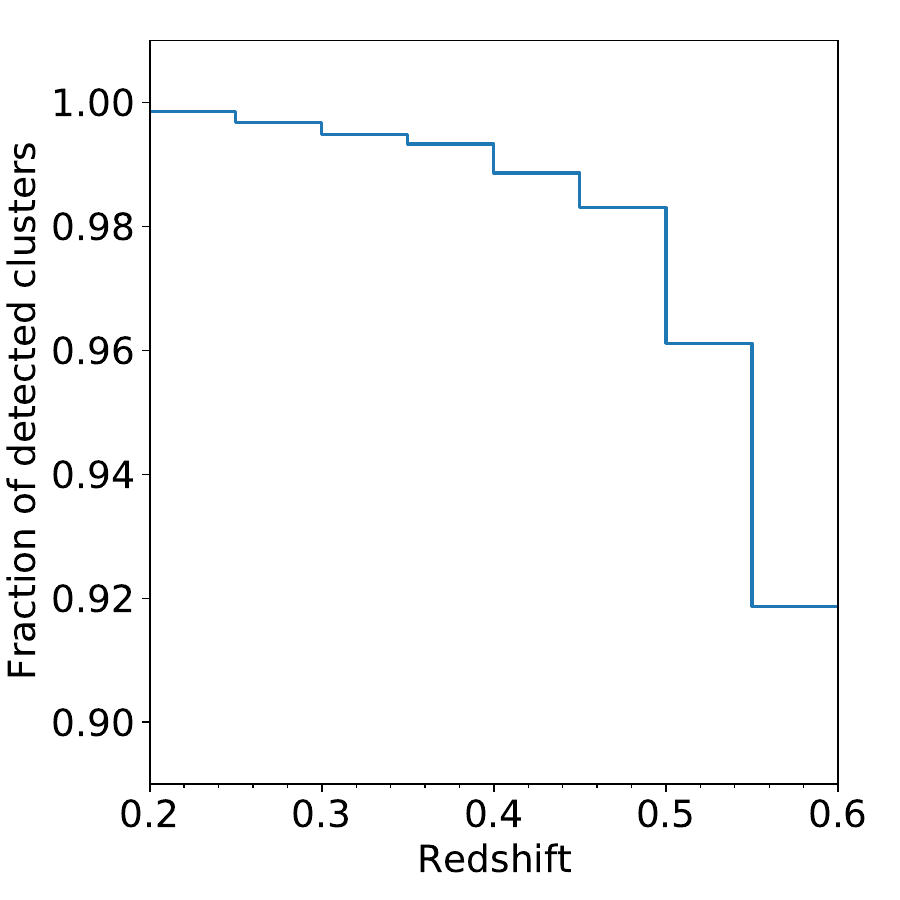}
    \caption{ The \YoCL completeness with respect to the redMaPPer cluster catalog as a function of redshift, when using the 512x512 (left) and the 1024x1024 (right) resampled SDSS images. When using 512x512 and 1024x1024  resampled SDSS images, \YoCL reaches a completeness of $\gtrsim 98\%$ for $z \gtrsim 0.3$ and $z \lesssim 0.4$, respectively.  In the other redshift ranges, the completeness is of $\sim 92-94 \%$. Overall, \YoCL has a very good performance in the redshift range covered by redMaPPer.} \label{fig:redshift}
\end{figure*}

\begin{figure*}[t!]
    \centering
    \includegraphics[width=0.35\textwidth]{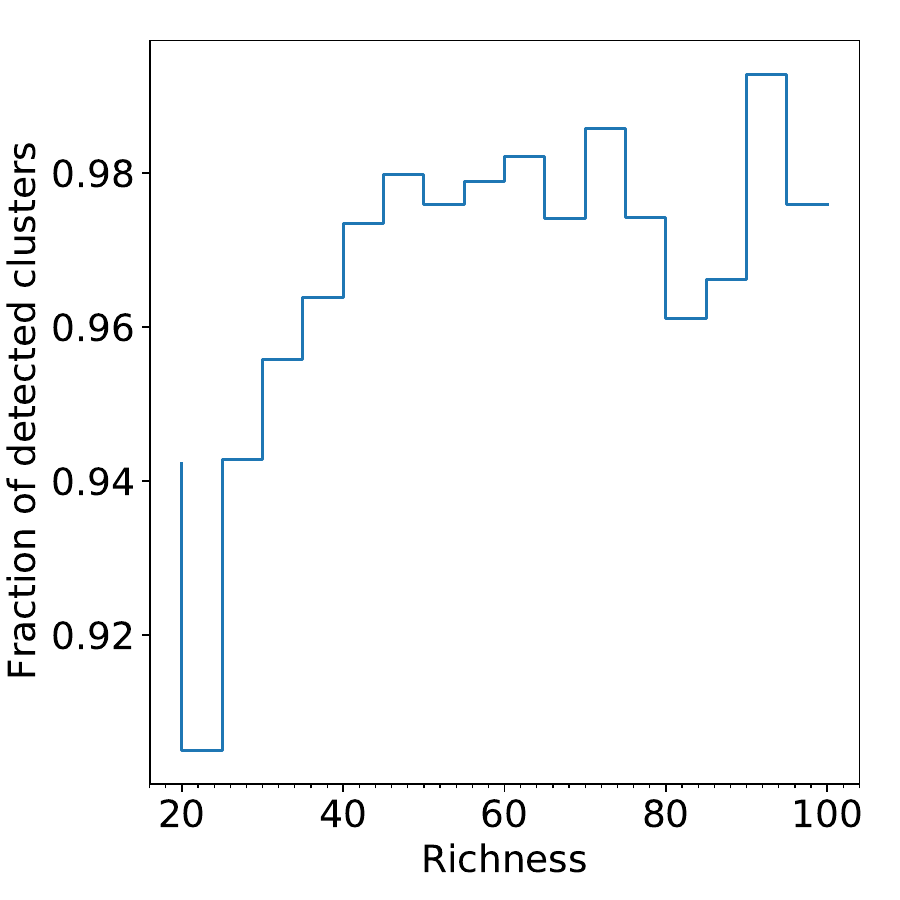}
     \includegraphics[width=0.35\textwidth]{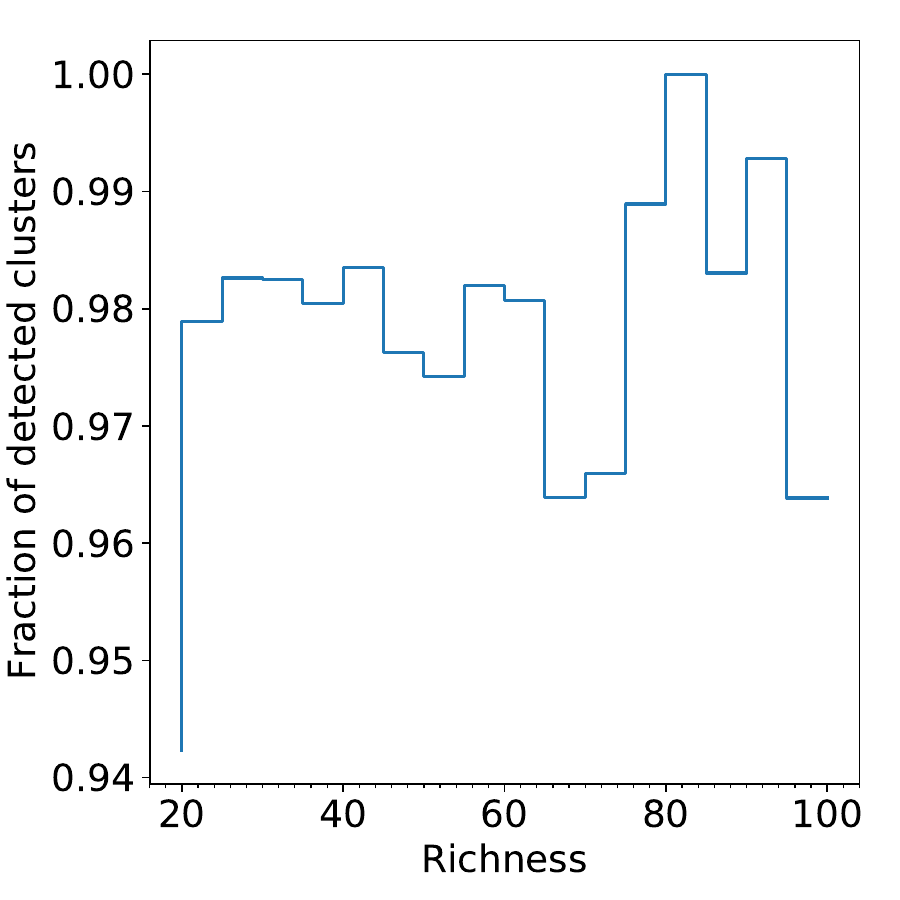}
    \caption{ The \YoCL completeness with respect to the redMaPPer cluster catalog as a function of the redMaPPer richness, when using the 512x512 (left) and the 1024x1024 (right) resampled SDSS images. When using 512x512 and 1024x1024 resampled SDSS images, \YoCL reaches a completeness of $\sim 98\%$ for $\lambda \gtrsim 40$. At lower richness the completeness is of $\sim 92-94 \%$.  Overall, \YoCL has a very good performance in the richness range covered by redMaPPer.} \label{fig:richness}
\end{figure*}

Then, we replaced the standard \Yo bounding box loss of Eq.~(\ref{eq:bbox_loss}) with the so-called generalized Intersection over Union (gIoU) loss of \citet{Rezatofighi_2018_CVPR}. In fact, the traditional IoU metric has among its main weaknesses the fact that it has a plateau (equal to 0) when the true and predicted bounding boxes are non-overlapping, making it impossible to optimize the corresponding loss term because of the vanishing gradient. The gIoU addresses this weakness by amending the IoU as follows:
\begin{equation}
    {\rm gIoU} = {\rm IoU} + \frac{\mathcal{U}}{\mathcal{A}_c} - 1 \,
\end{equation}
where $\mathcal{U}$ and $\mathcal{A}_c$ are respectively the areas of the union of the two boxes and the smallest box enclosing both boxes. This allows the gIoU to extend smoothly into negative values for boxes that are disjointed, and to tend towards $-1$ for more and more distant boxes (as ${\rm IoU}=0$ and $\mathcal{U}<<\mathcal{A}_c$). The gIoU was shown to yield better performance for multiple object detection compared to the standard bounding box metrics.

\subsubsection{Hyperparameter optimization}

Our hyperparameters were tested for best performance in optimizing the completeness and purity of our final cluster catalog.  Completeness and purity are the two parameters that characterize a galaxy cluster catalog and the algorithm selection function as a function of the cluster redshift and physical properties (such as mass, richness, X-ray luminosity, etc). The catalog completeness quantifies the fraction of true clusters that are detected by the algorithm. The catalog purity quantifies the fraction of detected clusters that are true, instead of false detections. In ML literature, the completeness corresponds to the recall, and the purity to the precision.

The dimension of the first network layer sets an upper limit for the amount of information that is used by the network, while also defining the size and complexity of the architecture. We start with SDSS images with size 2048$\times$2048 pixels, which corresponds to $\sim$13.5 x 13.5 arcmin$^2$ images, and to twice a typical cluster virial radius of 1~Mpc at z=0.3, the SDSS average redshift.  We resize each image to the dimensions of this network first layer by average pooling. In order to explore a trade-off between performance and precision, we consider two different input layer sizes, namely 1024$\times$1024 and 512$\times$512 pixels. We keep the same stride parameters as in the original \YoV publication, namely 8, 16, and 32. In practice, this means that for a given image the detection is performed at three scales on three $S{\times}\,S$ grids of cells, where $S = N/8$, $N/16$ and $N/32$ ($N$ is the pixel size of the image, i.e. 512 or 1024 in our case). Additionally, the number of bounding boxes per cell B is set to 3.

Multiple detections of the same object are discarded by applying a gIoU threshold of 0.5, which is similar to the Intersection-Over-Union (IoU) threshold in the original \YoV publication \footnote{We recall that the IoU is used to quantify the accuracy of a predicted
bounding box compared to the “true" box, by computing the ratio of the area of their intersection to the area of their union. A value of 1 corresponds to a perfect agreement, while a value that tends towards 0 indicates increasingly disjointed boxes and/or significantly different sizes.}. Table~\ref{tab:settings} shows the settings used for the network training.

\begin{figure*}
    \centering
    \includegraphics[width=0.47\textwidth]{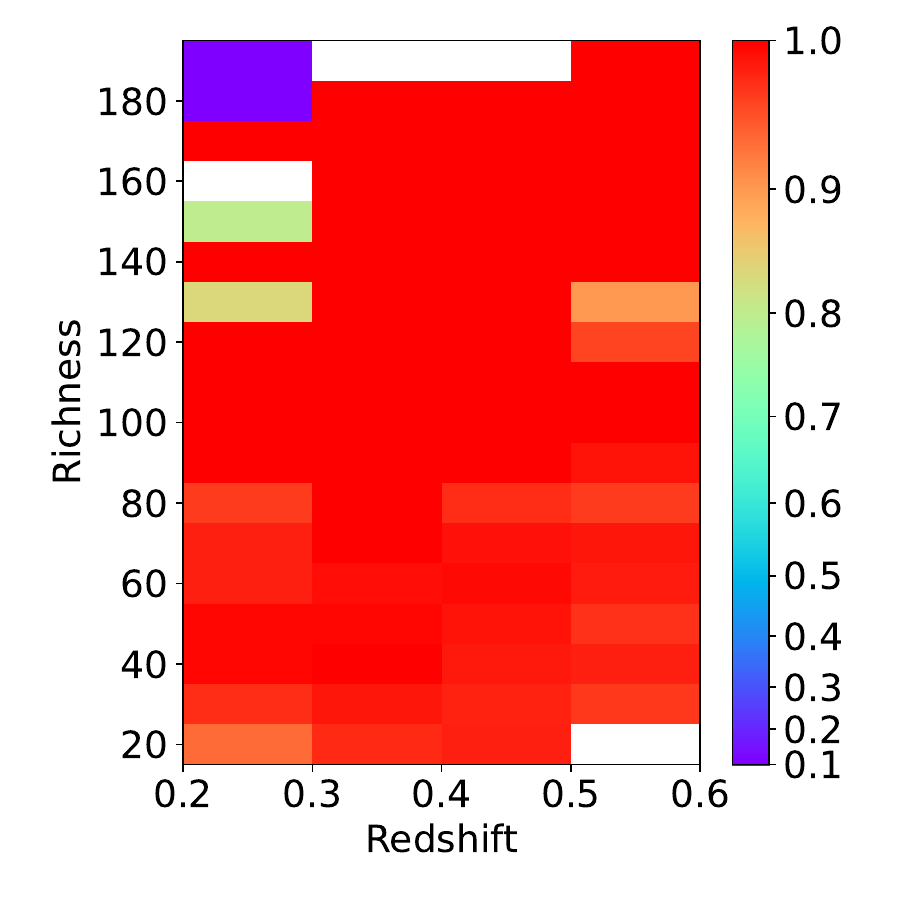}
     \includegraphics[width=0.47\textwidth]{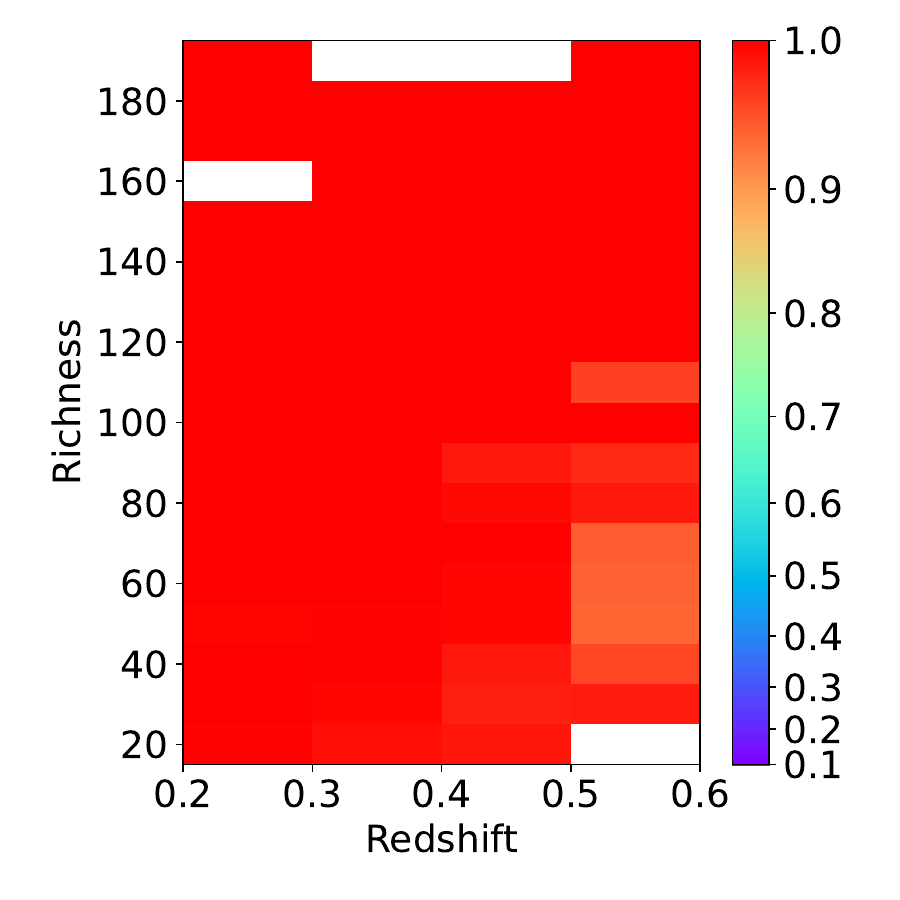}
      \caption{The \YoCL cluster sample completeness for redMaPPer detections for the images resampled to 512x512 (left) and to 1024x1024 (right) as a function of both richness and redshift. This figure synthesizes the conclusions of Fig.~\ref{fig:cp} and  Fig.~\ref{fig:redshift} as a function of both variables. On the right of each figure is the completeness scale.  }
      \label{fig:compl_zlam_2d}
\end{figure*}

\begin{figure}[t!]
    \centering
    \includegraphics[width=\hsize]{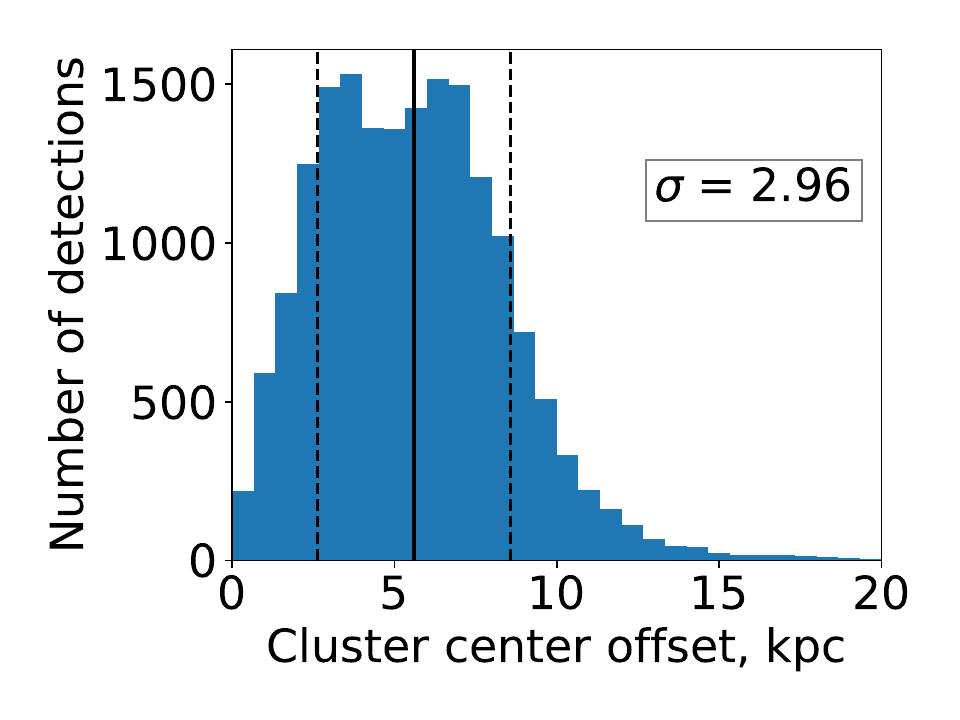}\\
    \includegraphics[width=\hsize]{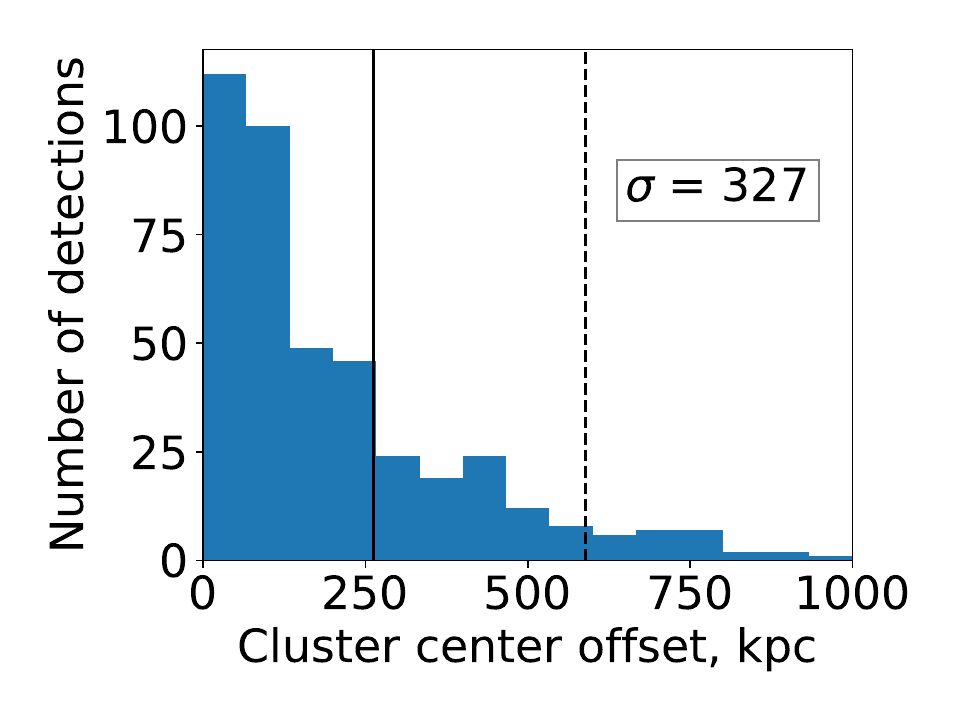}

    \caption{The distribution of the angular distance between cluster centers detected by \YoCL and redMaPPer (top panel), and \YoCL and the MCXC2021 clusters (bottom panel). } \label{fig:decenter}
\end{figure}

\subsection{Training and validation}

We trained \YoCL with about half ($\sim$12,000) of our selected redMaPPer cluster images and the same number of random SDSS blank field images of the same size. Our validation sample consists of the remaining cluster images and an equivalent number of blank field images. We calculate the validation loss at the end of each training epoch. We start by setting a learning rate of 10$^{-8}$, which grows slowly to 10$^{-4}$ during the four warm-up epochs, and then slowly decreases to 10$^{-6}$. 

Figure~\ref{fig:loss} shows the loss function for the training and validation sets for initial input images of size 512$\times$512 pixels and 1024$\times$1024 pixels. In both cases, there is a good agreement between the training and the validation loss functions, excluding significant overfitting and confirming our network stability.

The network output is a catalog of detection positions, bounding boxes, and their probability to be a cluster (hereafter detection probability). To build a sample of detected cluster candidates, we apply a probability cut by choosing only detections with a probability higher that a given threshold. To define which detection probability threshold to apply for our final cluster sample, we evaluate the performance of our network with respect to our \YoCL catalog completeness and purity. In this optimization, the redMaPPer cluster catalog is the truth.

Fig.~\ref{fig:cp} shows the completeness and purity of the detected cluster candidate sample as a function of a given  detection threshold when using the 512x512 and the 1024x1024 resampled SDSS images. In both cases, we build our final detection catalog by choosing the threshold that optimizes at the same time both completeness and purity, and which corresponds to the intersection of the completeness and purity curves in the figure. When using the 512x512 and the 1024x1024 resampled SDSS images, we obtain a completeness and purity of 95\% and 98\%, respectively, at the optimal threshold of 93\% and 60\%, respectively.

Our \YoCL final detection catalog (hereafter cluster catalog, with the caveat that the final detections are not confirmed galaxy clusters but cluster candidates with a given probability to be a cluster) includes the detections obtained when applying the above optimal thresholds.

\begin{figure*}[t!]
    \centering
    \includegraphics[width=\hsize]{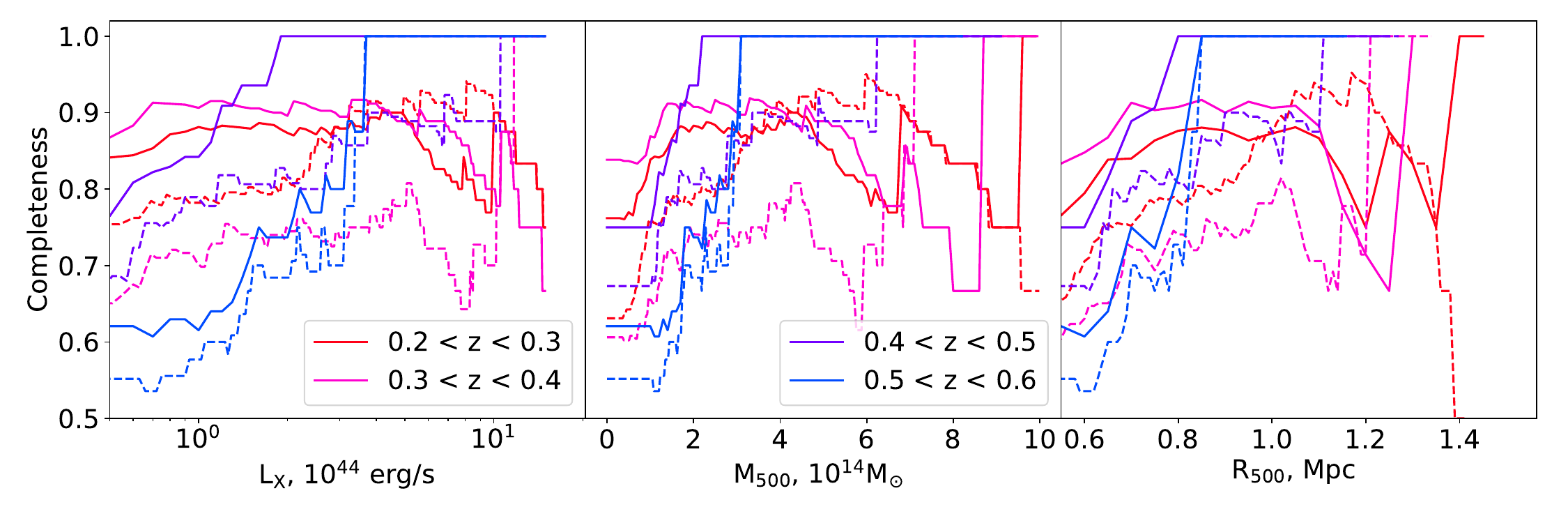}
    \caption{The \YoCL (continuous lines) and redMaPPer (dashed lines) cluster detection completeness above a given X-ray luminosity, $L_X$ (left panel), $M_{500}$ (middle panel) and R$_{500}$ (right panel).   \YoCL recovers all clusters at $L_X \gtrsim 1-3 \times 10^{44}$~erg/s, $M_{500} \gtrsim 2-3 \times 10^{14} M_{\odot}$, $R_{500} \gtrsim 0.75-0.8$~Mpc and $z \gtrsim 0.4$. At lower luminosity, mass, radius and redshift, its performance is worse. The redMaPPer algorithm recovers all clusters at at $L_X \gtrsim 3-9 \times 10^{44}$~erg/s, $M_{500} \gtrsim 2-6 \times 10^{14} M_{\odot}$, $R_{500} \gtrsim 0.8-1.2$~Mpc and $z \gtrsim 0.4$. At high redshifts both \YoCL and redMaPPer demonstrate similar performance that is limited by the SDSS depth.} \label{fig:mcxc2021cumcompl}
\end{figure*}

\begin{figure*}[t!]
    \centering
    \includegraphics[width=0.47\textwidth]{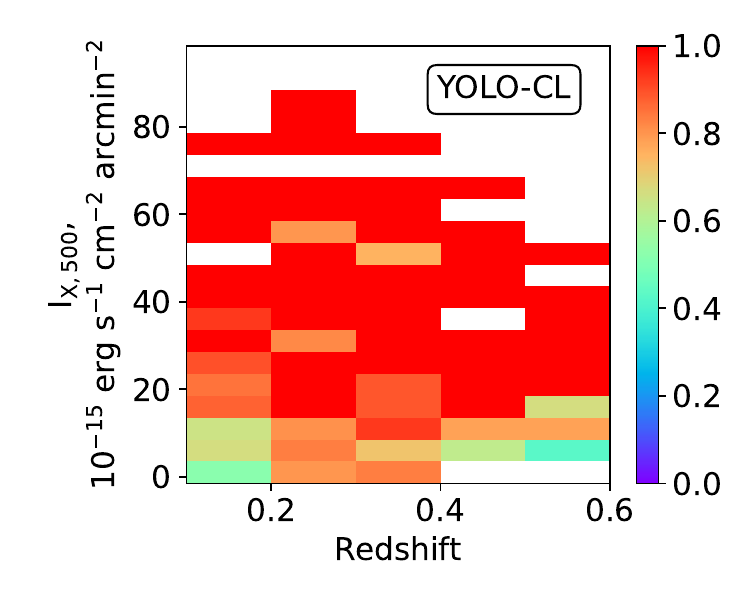}
     \includegraphics[width=0.47\textwidth]{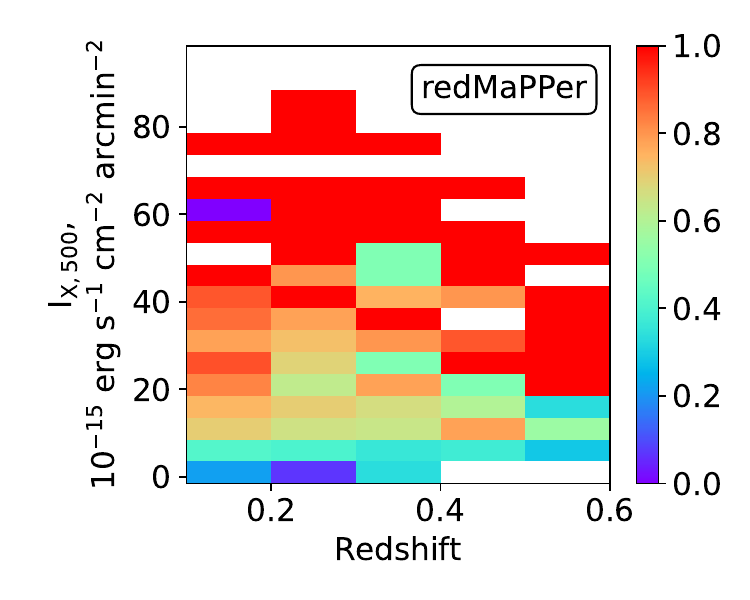}
    \caption{The \YoCL and redMaPPer MCXC2021 cluster detection completeness as a function of redshift and mean X-ray surface brightness. Left: \YoCL detects $\sim 98\%$ of the MCXC2021 clusters with $ {\rm I_{X, 500}} \gtrsim 20 \times 10^{-15} \ {\rm erg/s/cm^2/arcmin^2}$  at $0.2 \lesssim z \lesssim 0.6$  and $\sim 100\%$ of the MCXC2021 clusters with $ {\rm I_{X, 500}} \gtrsim 30 \times 10^{-15} \ {\rm erg/s/cm^2/arcmin^2}$ and $z \gtrsim 0.3$.  
Right: redMaPPer detects $\sim 98\%$ of the MCXC2021 clusters with $ {\rm I_{X, 500}} \gtrsim 55 \times 10^{-15} \ {\rm erg/s/cm^2/arcmin^2}$  at $0.2 \lesssim z \lesssim 0.6$  and $\sim 100\%$ of the MCXC2021 clusters with $ {\rm I_{X, 500}} \gtrsim 20 \times 10^{-15} \ {\rm erg/s/cm^2/arcmin^2}$ at  $0.5 \lesssim z \lesssim 0.6$.  On the right of each figure is the completeness scale. From this comparison, \YoCL is more complete than redMaPPer in detecting MCXC2021 clusters.}  \label{fig:mcxc2021compl}
\end{figure*}

\begin{figure*}[t!]
    \centering
    \includegraphics[width=\hsize]{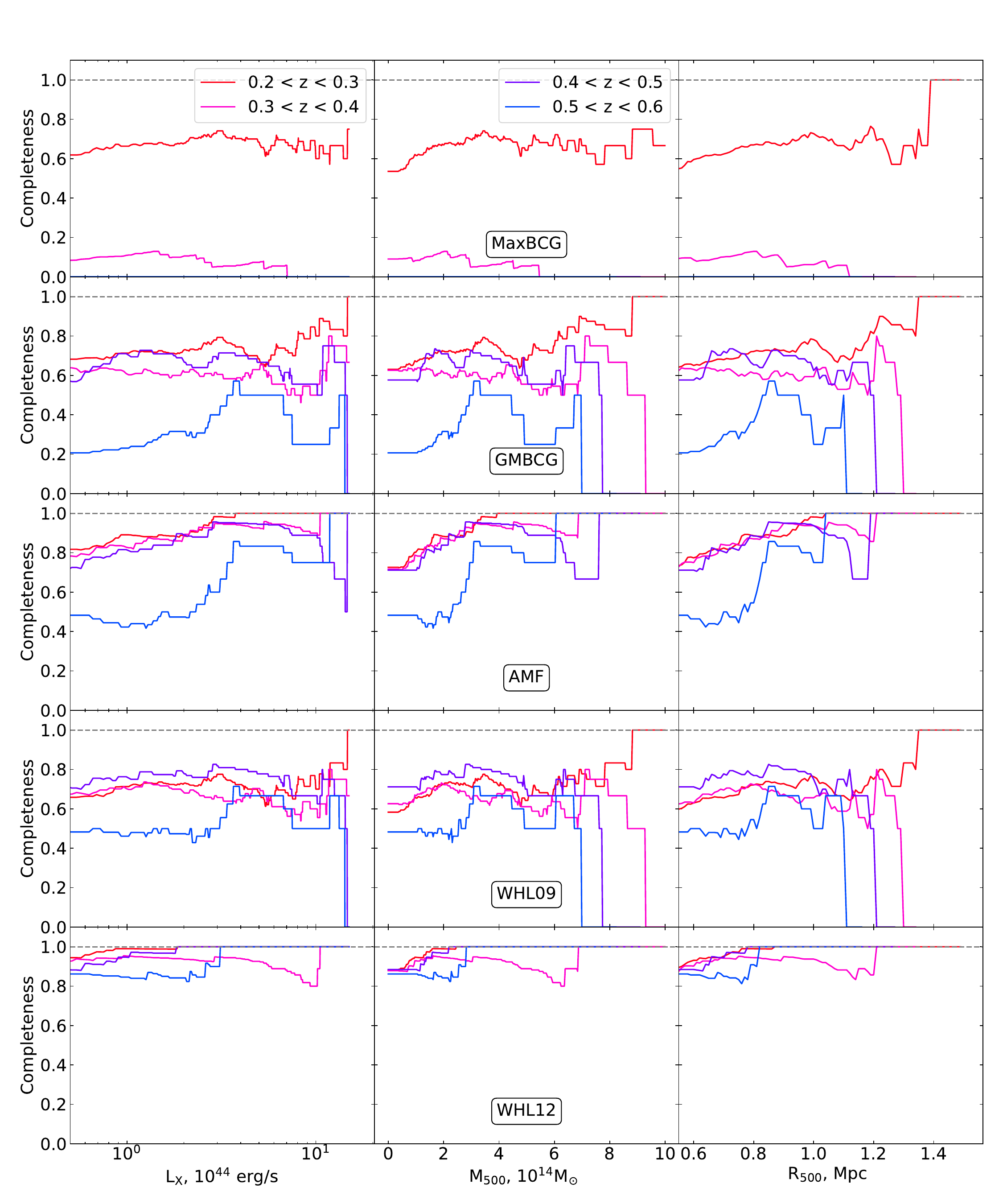}\\

    \caption{The fraction of MCXC2021 clusters recovered by tradition cluster detection methods in the SDSS (see text) from top to bottom: MaxBCG, GMBCG, AMF, WHL09 and WHL12. The details of cluster recovery in each case are detailed in the text. In all cases, except WHL12, their completeness is worse than that reached by redMaPPer and \YoCL. These results, compared with Fig.~8 outline the high performance of our \YoCL with respect to traditional cluster detection methods in the SDSS.}  \label{fig:othercats}
\end{figure*}

\section{Comparison to redMaPPer and X-ray cluster detections} \label{sec:red}

In this section, we compare our \YoCL cluster catalog  with our original redMaPPer catalog and X-ray cluster detections in the SDSS footprint.

\subsection{Comparison to redMaPPer detections}

Fig.~\ref{fig:redshift} and Fig.~\ref{fig:richness} show the fraction of redMaPPer clusters detected by \YoCL as a function of redshift and richness (i.e, the \YoCL completeness with respect to redMaPPer), respectively, and when using the 512x512 and the 1024x1024 resampled SDSS images. When using 512x512 resampled SDSS images, \YoCL detects $\sim 98\%$ of the redMaPPer clusters at $z \gtrsim 0.3$ and $\lambda \gtrsim 40$ (which corresponds to M$_{200} \sim 10^{14.3} h^{-1} M_{\odot}$ from \citealp{2017MNRAS.466.3103S}). For lower redshift ($0.2<z\lesssim 0.3$) and richness ($20< \lambda \lesssim 40$), the completeness is still very high at $\sim 92-93\%$. When using 1024x1024 resampled SDSS images, the performance of the network is similar as a function of richness, with a completeness of $\sim 98\%$, however the completeness as a function of redshift is different. At $z \lesssim 0.5$, \YoCL detects $>98\%$ of the redMaPPer clusters, and at higher redshift the completeness drops to $\sim 92\%$. This is synthesized in Fig.~\ref{fig:compl_zlam_2d} that shows the \YoCL sample completeness as a function of both redshift and richness.

Fig.~\ref{fig:decenter} shows the angular distance between the \YoCL cluster centers\footnote{defined as the center of the bounding box that hosts the cluster detection} and  redMaPPer cluster centers. The median angular distance between our and redMaPPer's cluster centers is of $\sim 5.6 \pm 2.9$~kpc, which is a very accurate recovery of redMaPPer cluster centers.

When checking if undetected clusters were found within detected cluster bounding boxes, we found that only $\sim 1\%$ of the redMaPPer cluster detections are not detected by \YoCL because they lie within the bounding box of another \YoCL detections (i.e., because they are superposed on the line of sight). This outlines the efficiency of the network in object separation. The remaining $99\%$ are detections either at $\lambda \lesssim 40$ or $z \gtrsim 0.5$, in the regime where redMaPPer detections are less complete because of the SDSS depth\citep{2014ApJ...783...80R}.

\subsection{Comparison to the MCXC2021 X-ray catalog } \label{sec:X-ray}

To validate the performance of \YoCL on a cluster sample independent of redMaPPer,  we  applied our network to galaxy clusters detected by X-ray emission and published in the MCXC2021 catalog\footnote{\url{https://www.galaxyclusterdb.eu/m2c/}}, the updated version of the MCXC catalog \citep{2011A&A...534A.109P}. X-ray detections confirm cluster detections as virialized dark matter haloes by their hot gas emission. We expect that this is not a mass selected sample because X-ray selected samples are biased towards relaxed cool core clusters \citeg{2016MNRAS.457.4515R}.

We built SDSS g, r and i-band color images for all MCXC2021 clusters in the SDSS footprint (927/1841 clusters), by generating 2048x2048 pixel (13.5x13.5 arcmin) cutouts with SkyServer\footnote{\url{http://skyserver.sdss.org/dr16/en/help/docs/api.aspx\#imgcutout}}, following the same procedure as for redMaPPer clusters. From this sample, we excluded the clusters that are only partially covered by the SDSS footprint.
As a comparison, we show the efficiency of redMaPPer on the same cluster sample. We crossmatched the MCXC2021 and redMaPPer catalogs within a range of 0.1 in redshift and a radius of 5~arcmin in position.

\begin{figure*}[t!]
    \centering
    \includegraphics[width=0.40\textwidth]{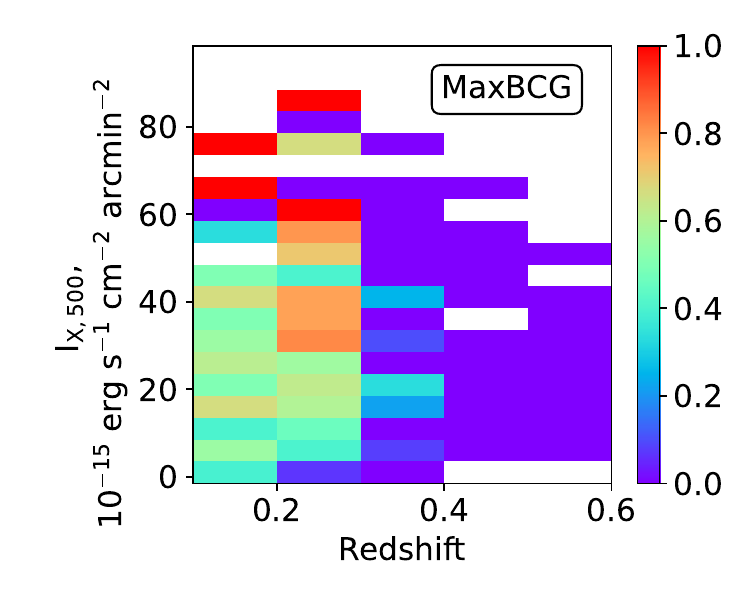}
    \includegraphics[width=0.40\textwidth]{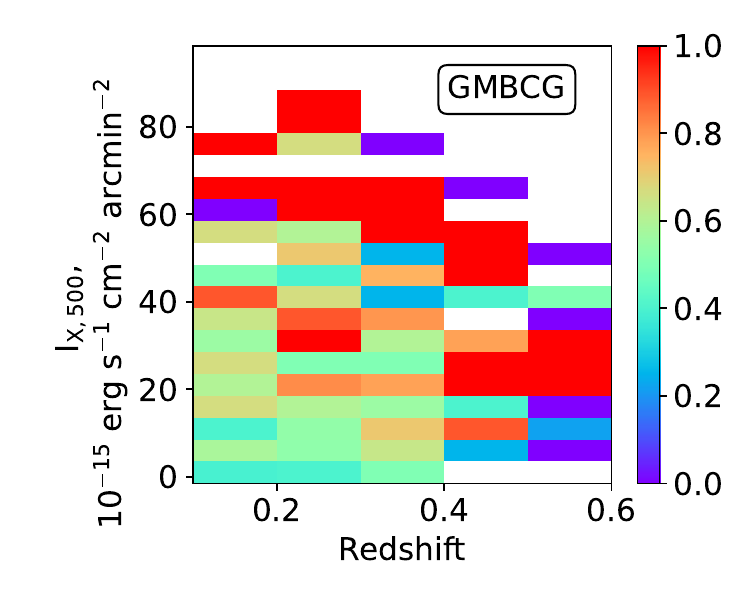}
    \includegraphics[width=0.40\textwidth]{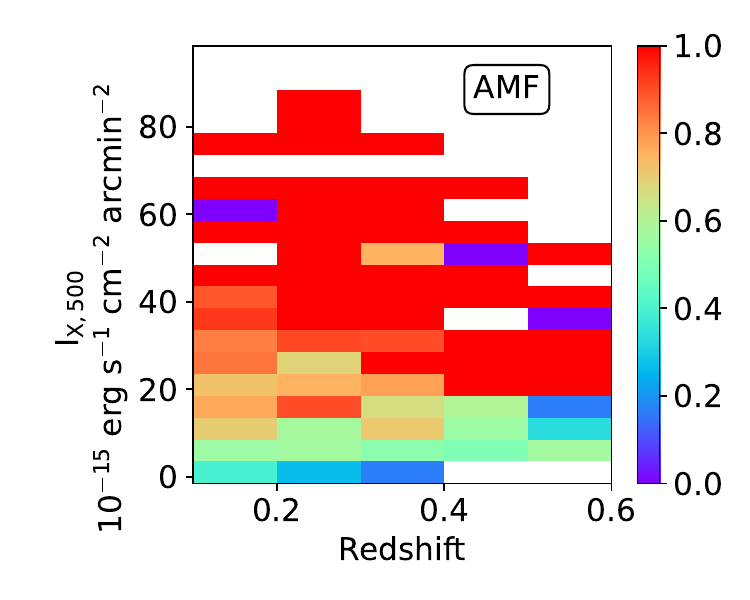}\\
    \includegraphics[width=0.40\textwidth]{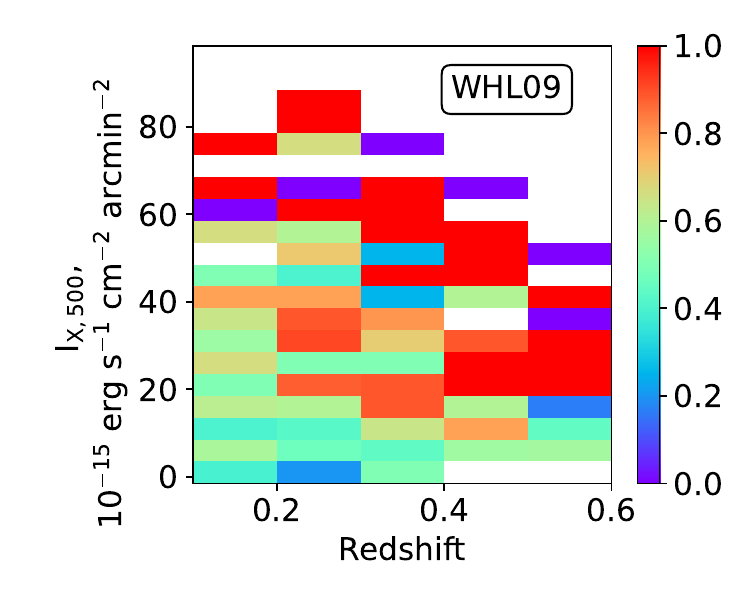}
    \includegraphics[width=0.40\textwidth]{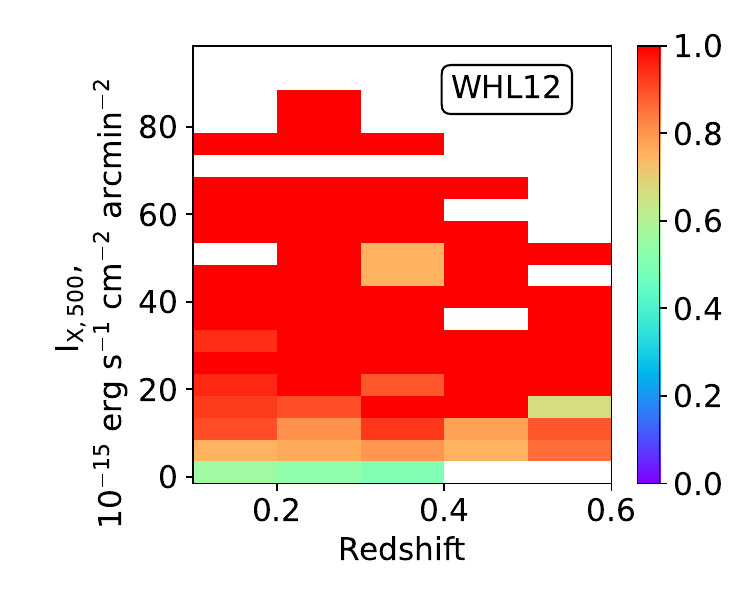}

    \caption{The completeness of the MaxBCG, GMBCG and AMF galaxy cluster catalogs as a function of redshift and mean X-ray surface brightness, $I_{X, 500}$. On the right of each figure is the completeness scale. All traditional cluster detection algorithms applied to SDSS are less complete than redMaPPer and \YoCL (see text), except AMF, which has a performance similar to redMapPPer, and WHL12, which has a performance similar to \YoCL. }  \label{fig:catsbxcompl}
\end{figure*}

Fig.~\ref{fig:mcxc2021compl} shows the \YoCL and redMaPPer cluster catalog completeness in redshift bins with respect to the MCXC2021 catalog as a function of X-ray luminosity $L_X$, derived cluster mass $M_{500}$ and radius $R_{500}$ \footnote{$M_{500}$ is defined as the mass within the circular region of radius $R_{500}$ containing a mean mass density equal to five hundred times the critical density of the Universe at a given redshift. The luminosity $L_X$ is the luminosity $L_{500}$ in the same region.}.
\YoCL recovers all clusters at $L_X \gtrsim 1-3 \times 10^{44}$~erg/s, $M_{500} \gtrsim 2-3 \times 10^{14} M_{\odot}$, $R_{500} \gtrsim 0.75-0.8$~Mpc and $z \gtrsim 0.4$. 
At lower luminosity, mass, radius and redshift, its performance is worse. redMaPPer recovers all clusters at at $L_X \gtrsim 3-9 \times 10^{44}$~erg/s, $M_{500} \gtrsim 2-6 \times 10^{14} M_{\odot}$, $R_{500} \gtrsim 0.8-1.2$~Mpc and $z \gtrsim 0.4$. 

To better compare \YoCL to redMaPPer, Fig.~\ref{fig:mcxc2021compl} shows the MCXC2021 cluster detection completeness as a function of the mean X-ray surface brightness ${\rm I_{X, 500}}$ \footnote{Defined as a mean X-ray flux within the region containing a mass density equal to five hundred times the critical density of the Universe at a given redshift divided by its angular circular area.} and redshift. The mean X-ray surface brightness  quantifies the average X-ray luminosity in a given area, and depends on the cluster luminosity and compactness, combining the information from the cluster $L_X$ and $R_{500}$ shown in Fig.~\ref{fig:mcxc2021compl}.

Our \YoCL network detects $\sim 98\%$ of the MCXC2021 clusters with $ {\rm I_{X, 500}} \gtrsim 20 \times 10^{-15} \ {\rm erg/s/cm^2/arcmin^2}$  at $0.2 \lesssim z \lesssim 0.6$  and $\sim 100\%$ of the MCXC2021 clusters with $ {\rm I_{X, 500}} \gtrsim 30 \times 10^{-15} \ {\rm erg/s/cm^2/arcmin^2}$ and $z \gtrsim 0.3$.  
redMaPPer detects $\sim 98\%$ of the MCXC2021 clusters with $ {\rm I_{X, 500}} \gtrsim 55 \times 10^{-15} \ {\rm erg/s/cm^2/arcmin^2}$  at $0.2 \lesssim z \lesssim 0.6$  and $\sim 100\%$ of the MCXC2021 clusters with $ {\rm I_{X, 500}} \gtrsim 20 \times 10^{-15} \ {\rm erg/s/cm^2/arcmin^2}$ at  $0.5 \lesssim z \lesssim 0.6$. 

 Fig.~\ref{fig:decenter} shows the angular distance between YOLO-CL and MCXC2021 cluster centers. The median angular distance is of $261 \pm 327$~kpc.  This large dispersion is consistent with the precision on the position of most of the MCXC2021 clusters that were detected by ROSAT with a $2.3$~arcmin angular resolution, which corresponds to $\sim 600$~kpc at the median MCXC2021 cluster redshift of $z \sim 0.3$, for our sample within the SDSS footprint.
 
From this comparison, \YoCL is more efficient than redMaPPer in detecting MCXC2021 clusters, which means that when optimizing \YoCL in terms of completeness and purity we improved cluster detection with respect to the redMaPPer algorithm, which provided our training sample. It is also interesting that the \YoCL selection function is approximately constant with redshift, with respect to the mean X-ray surface brightness.

\section{Discussion and Conclusions} \label{sec:discussion}

Our deep convolutional network \YoCL shows high completeness and purity in detecting galaxy clusters in the SDSS footprint. 
When compared to the existing redMaPPer catalog, we obtain cluster catalogs with completeness and purity of $95-98\%$ for our optimal thresholds when using 512x512 and the 1024x1024 resampled SDSS images. The X-ray parameter that we found more interesting is the mean X-ray surface brightness, which defines a clear threshold after which \YoCL and redMaPPer are $100\%$ complete. When comparing to the MCXC2021 X-ray detected clusters, \YoCL detects $\sim 98\%$ of the MCXC2021 clusters with $ {\rm I_{X, 500}} \gtrsim 20 \times 10^{-15} \ {\rm erg/s/cm^2/arcmin^2}$  at $0.2 \lesssim z \lesssim 0.6$  and $\sim 100\%$ of the MCXC2021 clusters with $ {\rm I_{X, 500}} \gtrsim 30 \times 10^{-15} \ {\rm erg/s/cm^2/arcmin^2}$ and $z \gtrsim 0.3$.  The lower detection rates for clusters at lower redshift could be explained by their large angular size that exceeds the size of image cutouts.

Several other SDSS cluster catalogs have been published, using different methods: the MaxBCG \citep{2007ApJ...660..239K}, WHL09/12 \citep{2009ApJS..183..197W, 2012ApJS..199...34W}, GMBCG \citep{2010ApJS..191..254H}, and AMF \citep{2011ApJ...736...21S} catalogs. We used the same methodology described for \YoCL and redMaPPer in Section~\ref{sec:X-ray} to asses the performance of these methods on the MCXC2021 cluster catalogs, and present our results in Fig.~\ref{fig:othercats} and Fig~\ref{fig:catsbxcompl}. MaxBCG recovers clusters with a $\sim$ 70\% completeness at $0.2<z<0.3$ in the entire range of $L_X$, $M_{500}$, and $R_{500}$ that we cover, with a $100\%$ recovery only for $R_{500} \gtrsim 1.4$~Mpc. At higher redshift, the completeness drops to $\lesssim 10\%$. GMBC recovers clusters with a  $\sim$100\% completeness only at $0.2<z<0.3$ and $L_X \gtrsim 10 \times 10^{44}$~erg/s, $M_{500} \gtrsim 8.5 \times 10^{14} M_{\odot}$, $R_{500} \gtrsim 1.3$~Mpc. At redshift $0.3<z<0.5$, its average completeness is of $50-70\%$. AMF recovers clusters with a  $\sim$ 100\% completeness at $0.2<z<0.3$ at $L_X \gtrsim 2 \times 10^{44}$~erg/s, $M_{500} \gtrsim 4 \times 10^{14} M_{\odot}$, $R_{500} \gtrsim 1$~Mpc and $0.2<z <0.3$. At higher redshift,  it recovers clusters with a  $\sim$ 100\%  at $L_X \gtrsim 10 \times 10^{44}$~erg/s, $M_{500} \gtrsim 6-8 \times 10^{14} M_{\odot}$, $R_{500} \gtrsim 1-1.3$~Mpc. WHL09 has a performance similar to AMF, and WHL12 is the most complete of those traditional methods, with a completeness of $\gtrsim 80\%$ for in the entire $L_X$, $M_{500}$, and $R_{500}$ ranges  at $0.2 \lesssim z \lesssim 0.6$ .  This last method reaches a completeness of $\sim 100\%$ for $L_X \gtrsim 0.8-3 \times 10^{44}$~erg/s, $M_{500} \gtrsim 2-3 \times 10^{14} M_{\odot}$, $R_{500} \gtrsim 0.8-0.9$~Mpc for $z<0.5$, and for  $L_X \gtrsim 10 \times 10^{44}$~erg/s, $M_{500} \gtrsim 7 \times 10^{14} M_{\odot}$, $R_{500} \gtrsim 1.2$~Mpc for $0.5<z<0.6$.

Fig~\ref{fig:catsbxcompl} summarize the performance of these three algorithms. While the MaxBCG, GMBCG, and WHL09 cluster catalogs are much less complete than the redMaPPer and \YoCL catalogs, the AMF cluster catalog results are very similar to redMaPPer's when considering the mean X-ray surface brightness. WHL12 shows a completeness very similar to \YoCL. The \YoCL performance is higher than most of the traditional detection algorithms applied to SDSS. Unfortunately, we can not complete our comparison using both completeness and purity, because the estimates of purity for each method are not homogeneous.

This confirms that the redMaPPer catalog that we use to train our network is very good in terms of recovery of X-ray cluster detections, and outlines the high performance of our \YoCL with respect to traditional cluster detection methods in the SDSS.

Together with its high performance in terms of completeness and purity, a strong advantage of galaxy cluster detection by deep learning networks is that clusters can be found without the need of measuring galaxy photometry and photometric redshifts. In fact, the direct use of color images allows us to skip the step of photometric and photometric redshift catalog preparation, and eliminates the systematic uncertainties that can be introduced during this process. This advantage has been pointed out in \citet{2019MNRAS.490.5770C}, which introduced for the first time the use of deep learning for cluster detection in SDSS with the development of Deep-CEE (Deep Learning for Galaxy Cluster Extraction and Evaluation). Deep-CEE is based on Faster region-based convolutional neural networks, trained on the  \citet{2012ApJS..199...34W}'s cluster catalog in the redshift range $0.05 \leq z < 0.8$.  As a proof of concept, they obtained completeness and purity of $\sim 75\%$ and $\sim 80\%$ when optimizing both on their validation sample and the redMaPPer catalog, respectively. 

The use of convolutional networks with color images also allows us to equally focus on the two main aspects of galaxy clusters that are used for detection: (i) the fact that they are galaxy overdensities, and (ii) the same distance/redshift, and therefore similar colors, of cluster members. The vast majority of existing detection algorithms focus on one of these aspects more than the another. Color images preserve both the information about galaxy positions and colors, without assumptions on the significance of the overdensity or galaxy colors.

We conclude that our \YoCL network has a higher performance in terms of completeness and purity in detecting galaxy clusters when compared to traditional cluster detection algorithms applied to SDSS images and catalogs. A strong advantage of deep learning networks is that clusters can be found without the need of measuring galaxy photometry and photometric redshifts, and the biases inherent to galaxy detection and these two measurements.

\section{Summary} \label{sec:summary}

We apply the \Yo object detection deep convolutional network to the detection of galaxy clusters in the SDSS survey. Our network implementation, \YoCL, is a modification of the original \YoV implementation to optimize galaxy cluster detection. 

\YoCL was trained and validated using three color images (in the g, r, i bandpasses) of 26,111 detections from redmappper cluster catalog and the equivalent number of SDSS blank field images. In the validation, we obtain an estimation of our network cluster catalog completeness and purity. To asses our sample completeness with respect to X-ray detected clusters, we compared our \YoCL detections to the MCXC2021 cluster catalog within the SDSS footprint.

\vspace{5mm}

Our results show that:

\begin{itemize}

\item When validated on the redMaPPer catalog, \YoCL detects $95\%$ and $98\%$ of the redMaPPer clusters when using 512x512 and 1024x1024 pixel resampled SDSS images, respectively. It reaches a purity of $95\%$ and $98\%$, calculated by applying the network to 512x512 and 1024x1024 pixel resampled SDSS blank fields.

\vspace{2mm}

\item When compared to the redMaPPer detection of the same X-ray detected MCXC2021 clusters, \YoCL is more complete at lower $L_X$,  $M_{500}$ and  $R_{500}$ than redMaPPer. This means that the neural network improved the cluster detection efficiency of its training sample. In fact, \YoCL recovers all clusters at $L_X \gtrsim 1-3 \times 10^{44}$~erg/s, $M_{500} \gtrsim 2-3 \times 10^{14} M_{\odot}$, $R_{500} \gtrsim 0.75-0.8$~Mpc and $z \gtrsim 0.4$. 
At lower luminosity, mass, radius and redshift, its performance degrades. In comparison, redMaPPer recovers all clusters at at $L_X \gtrsim 3-9 \times 10^{44}$~erg/s, $M_{500} \gtrsim 2-6 \times 10^{14} M_{\odot}$, $R_{500} \gtrsim 0.8-1.2$~Mpc and $z \gtrsim 0.4$.  

\vspace{2mm}

\item  \YoCL detects lower mean X-ray surface brightness $ {\rm I_{X, 500}}$ clusters with respect to redMaPPer. In fact, \YoCL detects $\sim 98\%$ of the MCXC2021 clusters with $ {\rm I_{X, 500}} \gtrsim 20 \times 10^{-15} \ {\rm erg/s/cm^2/arcmin^2}$  at $0.2 \lesssim z \lesssim 0.6$  and $\sim 100\%$ of the MCXC2021 clusters with $ {\rm I_{X, 500}} \gtrsim 30 \times 10^{-15} \ {\rm erg/s/cm^2/arcmin^2}$ and $z \gtrsim 0.3$, while redMaPPer detects $\sim 98\%$ of the MCXC2021 clusters with $ {\rm I_{X, 500}} \gtrsim 55 \times 10^{-15} \ {\rm erg/s/cm^2/arcmin^2}$  at $0.2 \lesssim z \lesssim 0.6$  and $\sim 100\%$ of the MCXC2021 clusters with $ {\rm I_{X, 500}} \gtrsim 20 \times 10^{-15} \ {\rm erg/s/cm^2/arcmin^2}$ at  $0.5 \lesssim z \lesssim 0.6$. 

 \vspace{2mm}
 
 \item The \YoCL selection function is approximately constant with redshift, with respect to the MCXC2021  cluster mean  X-ray surface brightness.

\vspace{2mm}

\item When comparing to other traditional detection algorithms applied to the SDSS survey, we confirm that redMaPPer is an excellent choice to train our network, in terms of recovery of X-ray cluster detections. This comparison also outlines the high performance of our \YoCL with respect to most of the traditional cluster detection methods in the SDSS.

\end{itemize}

Our results show that our \YoCL  cluster sample has a very high completeness when compared to redMaPPer and other traditional cluster detection algorithms in detecting the X-ray detected MCXC2021 clusters. \YoCL also shows a very high level of purity, measured using SDSS blank field images. As pointed out in the first implementation of a deep convolutional network for galaxy cluster detection \citep{2019MNRAS.490.5770C},   deep learning networks  have a strong advantage of galaxy cluster detection with respect to traditional techniques because they do not need galaxy photometric and photometric redshift catalogs. This eliminates the systematic uncertainties that can be introduced during the source detection, and the measurements of photometry  and photometric measurements, and focuses the detection method on the two main aspects of galaxy cluster detection, the search for overdensities of galaxies that have similar colors because they are at the same redshift, without assumptions on the significance of the overdensity or galaxy colors. Our results highlight another advantage: a higher cluster catalog completeness than traditional cluster detection algorithms applied to the SDSS and a very high purity. Interestingly, the \YoCL selection function is approximately constant with redshift, with respect to the mean X-ray surface brightness.

We conclude that deep convolutional network for galaxy cluster detection are an efficient alternative to traditional cluster detection methods, and it is worth exploring their performance for future cosmological cluster catalogs for large-scale surveys, such as the Rubin/LSST, Euclid and the Roman Space Telescope. 

\begin{acknowledgements}
We describe below the author's contributions. In 2019-2021, SI modified the original YOLO network to adapt it for galaxy cluster detection. He conceived YOLO-CL with SM and developed the network and analysis. He is the main YOLO-CL developer. He applied YOLO-CL to SDSS images that he produced as described in the paper, and produced the plots for the completeness and purity analysis. He was the main author of the text describing YOLO-CL. KS joined our team in 2021, reproduced Ilic's work and compared the final YOLO-CL cluster catalog to the redMapper catalog and other cluster finders. SM conceived the YOLO-CL network with SI, supervised the work of KS and SI and was the main writer of the paper text, the contact with the editor and answered the referee report.
We thank Universit\'e Paris Cit/'e (UPC), which funded KG's Ph.D. research, and University Paris Science \& Lettres (PSL), which funded SI postdoctoral research. We thank our PSL and UPC colleagues, and collaborators from the LightOn (https://lighton.ai/) company and the \'Ecole Normale Superieure (ENS), Laurent Daudet, Florent Krzakala, and Am\'elie Chatelain, for fruitful discussions. We thank Jean-Baptiste Melin and James Bartlett for useful discussions and help in choosing the X-ray catalog that we use in this paper. We gratefully acknowledge support from the CNRS/IN2P3 Computing Center (Lyon - France) for providing computing and data-processing resources needed for this work. This research has made use of the M2C Galaxy Cluster Database, constructed as part of the ERC project M2C (The Most Massive Clusters across cosmic time, ERC-Adv grant No. 340519).  This work was
supported by the French Space Agency (CNES).
\end{acknowledgements} 
\clearpage

\bibliographystyle{aa}
\bibliography{main.bib}

\end{document}